\documentclass{article}
\usepackage{graphicx} % Required for inserting images
\usepackage{booktabs}
\usepackage{multicol, multirow}
\usepackage{float}
\usepackage{verbatim}
\usepackage{cite}
\usepackage{authblk}
\usepackage{amsmath,amsfonts}
\usepackage{booktabs}
\usepackage{stfloats}
\usepackage{url}
\usepackage{rotating}
\usepackage{geometry}
\usepackage{booktabs}
\usepackage{multicol, multirow}
\usepackage{url}

\title{\textit{TempEE}: Temporal-Spatial Parallel Transformer for Radar Echo Extrapolation Beyond Auto-Regression}
\author[a,b]{Shengchao Chen\footnote{Work done during an internship at Guangdong-Hongkong-Macao Greater Bay Area Weather Research Center for Monitoring Warning and Forecasting (Shenzhen Institute of Meteorological Innovation). Emali: pavelchen@ieee.org.}}
\author[a]{Ting Shu\thanks{Corresponding author: shuting@gbamwf.com}}
\author[c]{Huan Zhao}
\author[d]{Guo Zhong}
\author[e]{Xunlai Chen}

\affil[a]{Guangdong-Hongkong-Macao Greater Bay Area Weather Research Center for Monitoring Warning and Forecasting (Shenzhen Institute of Meteorological Innovation), Shenzhen, China}
\affil[b]{Australian Artificial Intelligence Institute, School of Computer Science, FEIT, University of Technology Sydney, Sydney, NSW, Australia}
\affil[c]{The Chinese University of Hong Kong, Shenzhen, China}
\affil[d]{School of Information Science and Technology, Guangdong University of Foreign Studies, Guangzhou, China}
\affil[e]{Shenzhen Meteorological Bureau / Shenzhen Key laboratory of severe weather in south China, Shenzhen, China}
\date{The manuscript have been accepted by IEEE Transactions on Geoscience and Remote Sensing\footnote{https://doi.org/10.1109/TGRS.2023.3311510}}

\begin{document}

\maketitle

\begin{abstract}
Meteorological radar reflectivity data (i.e. radar echo) significantly influences precipitation prediction. It can facilitate accurate and expeditious forecasting of short-term heavy rainfall bypassing the need for complex Numerical Weather Prediction (NWP) models. In comparison to conventional models, Deep Learning (DL)-based radar echo extrapolation algorithms exhibit higher effectiveness and efficiency. Nevertheless, the development of reliable and generalized echo extrapolation algorithm is impeded by three primary challenges: cumulative error spreading, imprecise representation of sparsely distributed echoes, and inaccurate description of non-stationary motion processes. To tackle these challenges, this paper proposes a novel radar echo extrapolation algorithm called Temporal-Spatial Parallel Transformer, referred to as \textit{TempEE}. \textit{TempEE} avoids using auto-regression and instead employs a one-step forward strategy to prevent cumulative error spreading during the extrapolation process. Additionally, we propose the incorporation of a Multi-level Temporal-Spatial Attention mechanism to improve the algorithm's capability of capturing both global and local information while emphasizing task-related regions, including sparse echo representations, in an efficient manner. Furthermore, the algorithm extracts spatio-temporal representations from continuous echo images using a parallel encoder to model the non-stationary motion process for echo extrapolation. The superiority of our \textit{TempEE} has been demonstrated in the context of the classic radar echo extrapolation task, utilizing a real-world dataset. Extensive experiments have further validated the efficacy and indispensability of various components within \textit{TempEE}.
\end{abstract}
\section{Introduction}
\label{sec_int}
Extreme meteorological events, such as typhoons and thunderstorms, can cause heavy rainfall within local areas in a short period. These events can lead to life-threatening accidents, such as flash floods, mud-rock flows, urban waterlogging, and landslides, and can cause significant damage to infrastructure~\cite{qiu2016prediction}. Hence, accurate, reliable, and timely precipitation forecasting is essential to mitigate the effects of extreme weather. While Numerical Weather Prediction (NWP) models~\cite{zhao2018one, grams2018atmospheric} can account for a wide range of atmospheric physical constraints and explain some long-term weather processes, their uncertainties, high computational cost, and significant storage resource limite their effectiveness for short-term precipitation forecasting~\cite{chen2023spatial,chen2023prompt}. Therefore, alternative methods that provide more stable, real-time, and cost-effective short-term precipitation forecasts should be explored~\cite{ramirez2005artificial}.

The predicting of future continuous information about precipitation cloud movement through meteorological radar echos extrapolation is crucial for short-term precipitation forecasting. However, traditional radar echo extrapolation methods~\cite{li2004applications,wong2009towards,woo2017operational,germann2002scale,germann2004scale}, such as using motion vectors to model changes in echo patterns~\cite{li2004applications}, have limited accuracy in modeling nonlinear variations and uncertainties in complex weather systems. Artificial Intelligence (AI), particularly Deep Learning (DL), have provided alternative methods \cite{yang2023self,fang2023stunner,shi2015convolutional,hu2021towards,xie2020energy,sun2021mfbcnnc,wang2017predrnn,wang2018predrnn++,wang2019memory,chai2022cms} for radar echo extrapolation. These methods employ temporal-spatial forecasting, where the model learns temporal-spatial information from historical observations to predict future echo motion.

DL-based strategies for radar extrapolation can be categorized into two types: those based on Convolutional Neural Networks (CNN)~\cite{ayzel2019all,agrawal2019machine,fang2023stunner,song2019deep,castro2021stconvs2s} and those based on Recurrent Neural Networks (RNN)~\cite{shi2015convolutional,wang2017predrnn,wang2018predrnn++,wang2019memory,chai2022cms,sonderby2020metnet,klocek2021ms,jing2020hprnn,9837952}. The former is better at modeling spatial representation of radar echos, while RNN-based is superior in capturing temporal correlations among various motion processes of radar echoes. To overcome the limitations of these models, recent studies have proposed hybrid models~\cite{shi2015convolutional,chai2022cms,fang2023stunner} that combine CNN and RNN. However, these models are not without limitations, such as being vulnerable to weak sparse echo features and non-stationary weather processes. Additionally, as extrapolation time increases, these models may suffer from a rapid decay of echo refinement and prediction accuracy, which could potentially impact their reliability and availability.

To address these limitations, this paper proposes a novel DL model called the \textbf{Tem}poral-Spatial \textbf{P}arallel Transformer for Radar \textbf{E}cho \textbf{E}xtrapolation (dubbed \textbf{\textit{TempEE}}). The \textit{TempEE} constructs a comprehensive representation of the complex temporal-spatial relationships between historical observations and future echoes for accurate and timely precipitation forecasting. Compared to the auto-regression-based extrapolation method, \textit{TempEE} uses a one-step forward strategy to learn the distribution of radar echoes and motion trends in parallel, resulting in an advanced temporal and spatial representation. Furthermore, a new attentional fraction calculation strategy is proposed in the overall framework to represent the global-local echo distribution with low computational effort. \textit{TempEE} is an innovative approach to improving the accuracy of precipitation forecasting based on radar echo image extrapolation tasks.

The contributions of this work are summarized in four-fold: 

\begin{itemize}
    \item A Transformer-based model is proposed for precipitation forecasting, which reliably predicts complex radar echo distributions and motion trends, including dense and sparse distributions as well as stationary and non-stationary motion trends. The model outperforms auto-regression-based extrapolation algorithms in terms of effectiveness.
    \item To flexibly and effectively capture non-stationary motion trends and sparse radar echo distributions, we propose a parallel encoder structure comprising a Temporal Encoder (TE) and a Spatial Encoder (SE). The TE creates temporal correlations between echo distributions, while the SE deals with the spatial representation of echo distributions.
    \item We introduce a Temporal-Spatial Decoder (TSD) based on reverse random sampling to enhance the ability to extract global-local information and spatiotemporal correlations. During the training phase, this approach randomly provides prior knowledge, which prompts the model to learn a complete representation of the trend. Additionally, we adopt a Multi-level Spatio-Temporal Attention (MSTA) to refine echo features in a lightweight manner with low computational complexity.
    \item We train and test the proposed model on a real-world radar echo image dataset and perform an extensive study to analyze the contribution of each component. The experimental results demonstrate that the model accurately predicts both stationary and non-stationary weather processes and significantly outperforms other DL-based models in terms of prediction accuracy and efficiency.
\end{itemize}

The remainder of this work is in the following: Section \ref{sec_rel} reviews related works. Section \ref{sec_met} describes the proposed model in detail. The the specific implementation details of the experiment and the results are given in Section \ref{sec_exp}. The last section (Section \ref{sec_con}) concludes this work.

\section{Related work}
\label{sec_rel}
This section summarizes the relevant representative approaches on this work, including traditional radar echo extrapolation methods, DL-based radar echo extrapolation, and vision Transformers (ViTs).

\subsection{Traditional Radar Echo Extrapolation Methods}
The conventional numerical echo extrapolation models require determining the motion region. The Tracking Radar Echoes Correlation (TREC) algorithm~\cite{li2004applications} analyzes the correlation of chunks in the divided echo image to predict echoes. Optical-flow~\cite{wong2009towards, woo2017operational} have been widely utilized for precipitation nowcasting via establishing spatio-temporal relationships between adjacent frames to capture motion information. Multi-scale optical flow analysis of variance (MOVA)~\cite{wong2009towards} applies multiple scale constraints to stationary the motion field and reduce the extrapolation discontinuity. Real-time Optical flow by Variational methods for Echoes of Radar (ROVER)~\cite{woo2017operational} enhances the continuity by preprocessing the radar reflectance of aquatic animals before applying MOVA. Semi-Lagrangian advection can be used for future extrapolation after implementing optical flow~\cite{germann2002scale, germann2004scale}.

Numerical models have a drawback in that they estimate the motion of echoes based on only a few observations, which makes it difficult to establish the temporal correlation among them~\cite{imhoff2022large,marrocu2020performance}. Optical flow-based approaches rely on the assumption of constant brightness, which is easily violated in the motion pattern of radar echoes~\cite{sun2008learning, novak2009quantitative}. The assumption of no growth or decay of precipitation during semi-Lagrangian extrapolation (known as Lagrangian persistence) is inaccurate~\cite{han2019convolutional,han2022toward,gong2023enhancing}.

\subsection{DL-based Radar Echo Extrapolation}
Numerous deep learning (DL)-based models have been developed for spatiotemporal forecasting of radar echoes. These data-driven techniques use powerful nonlinear fitting units to abstract complex spatiotemporal representations among echo motion processes without considering potential physical constraints in the atmosphere~\cite{luo2022experimental,dai2022mstcgan,bai2022rainformer,9837952}. The two typical strategies for echo extrapolation are the convolutional neural network (CNN)-based and recurrent neural network (RNN)-based approaches.

CNN-based approaches simultaneously process spatial and temporal features in the echo sequence by using convolution kernels to directly learn the translation mode from historical observations to future representation. Ayzel et al.~\cite{ayzel2019all} used an all-CNN model that performs comparably to the optical flow algorithm. Agrawal et al.~\cite{agrawal2019machine} employed a U-Net \cite{ronneberger2015u} that fuses features from abstract to concrete levels to predict echoes, outperforming optical flow, persistence, and high-resolution rapid refresh (HRRR) numerical models in 1-hour forecasting. Song et al.~\cite{song2019deep} proposed a modified U-Net based on ResNet~\cite{he2016deep} and attention module~\cite{woo2018cbam} to provide a more accurate representation of echo details. Li et al. \cite{li2021msdm} incorporated satellite brightness as auxiliary information inputs to help predict echo intensity using a U-Net model. Castro et al.~\cite{castro2021stconvs2s} proposed a 3D convolutional layer to address spatiotemporal causal constraints in weather forecasting.

The RNN-based approach uses recurrent connections to establish temporal correlations among echo motion processes. Shi et al.~\cite{shi2015convolutional} introduced a novel long short-term memory (LSTM) with conventional operation to model echo motion patterns for spatially adjacent regions. Wang et al. \cite{wang2019memory} proposed a non-stationary learning unit to improve the learning and perception of long-term echo variations. Sønderby et al.~\cite{sonderby2020metnet} built a model for predicting high-resolution precipitation using ConvLSTM relying on multiple data sources, including echo intensity, topography, and satellite images. Klocek et al. \cite{klocek2021ms} combined ConvLSTM with a numerical model to improve the forecasting accuracy of long-term participation. Jing et al.~\cite{jing2020hprnn} used a coarse-to-fine recurrent strategy to improve RNN models' long-term inference performance.

Although CNN- and RNN-based approaches are useful in learning limited representations of echoes from historical observations and period-limited temporal correlations, they have limitations. These methods are vulnerable to weak sparse echo features and non-stationary weather processes, which may affect their reliability~\cite{fang2023stunner}. Additionally, the rapid decay of echo refinement and accuracy with increasing extrapolation time is a significant drawback of run-based methods, which may limit their availability~\cite{9837952}.

\subsection{Vision Transformers (ViTs)}
The Transformer revolutionized the field of Natural Language Processing (NLP) by employing self-attention to effectively handle long-term dependencies in sequential data~\cite{vaswani2017attention}. This achievement inspired researchers to propose Vision Transformers (ViTs)~\cite{dosovitskiy2020image, chen2023interpretable} as a technique for encoding image token sequences. However, to enhance the applicability and performance of Transformer in image domains, several approaches have been proposed, some of which come at the cost of increased computational complexity. Dosovitskiy et al.~\cite{dosovitskiy2020image} used patch segmentation to divide the original image into several small patches and extract tokens separately, while Touvron et al.~\cite{touvron2021going} improved the attention mechanism by introducing formaldehyde and class-specific attention. Similarly, Fan et al.~\cite{fan2021multiscale} devised a multiscale ViT structure, and Zhou et al.~\cite{zhou2021deepvit} regenerated attention maps to promote diversity at various levels of representation. In this paper, we introduce an efficient method for attention computation with low complexity. Our method captures both global-local relations among the input feature maps within the ViT. Our goal is to achieve state-of-the-art performance while ensuring computational efficiency.

% In this paper, we present a computational method for attention computation that is efficient and has low computational complexity.
% Our proposed method models the global and local relationships of the input feature maps in the ViT. Our goal is to achieve state-of-the-art performance while keeping computational costs to a minimum.

\section{Temporal-Spatial Parallel Transformer}
\label{sec_met}
\begin{figure*}[tbh]
\centering
\includegraphics[width=0.85\textwidth]{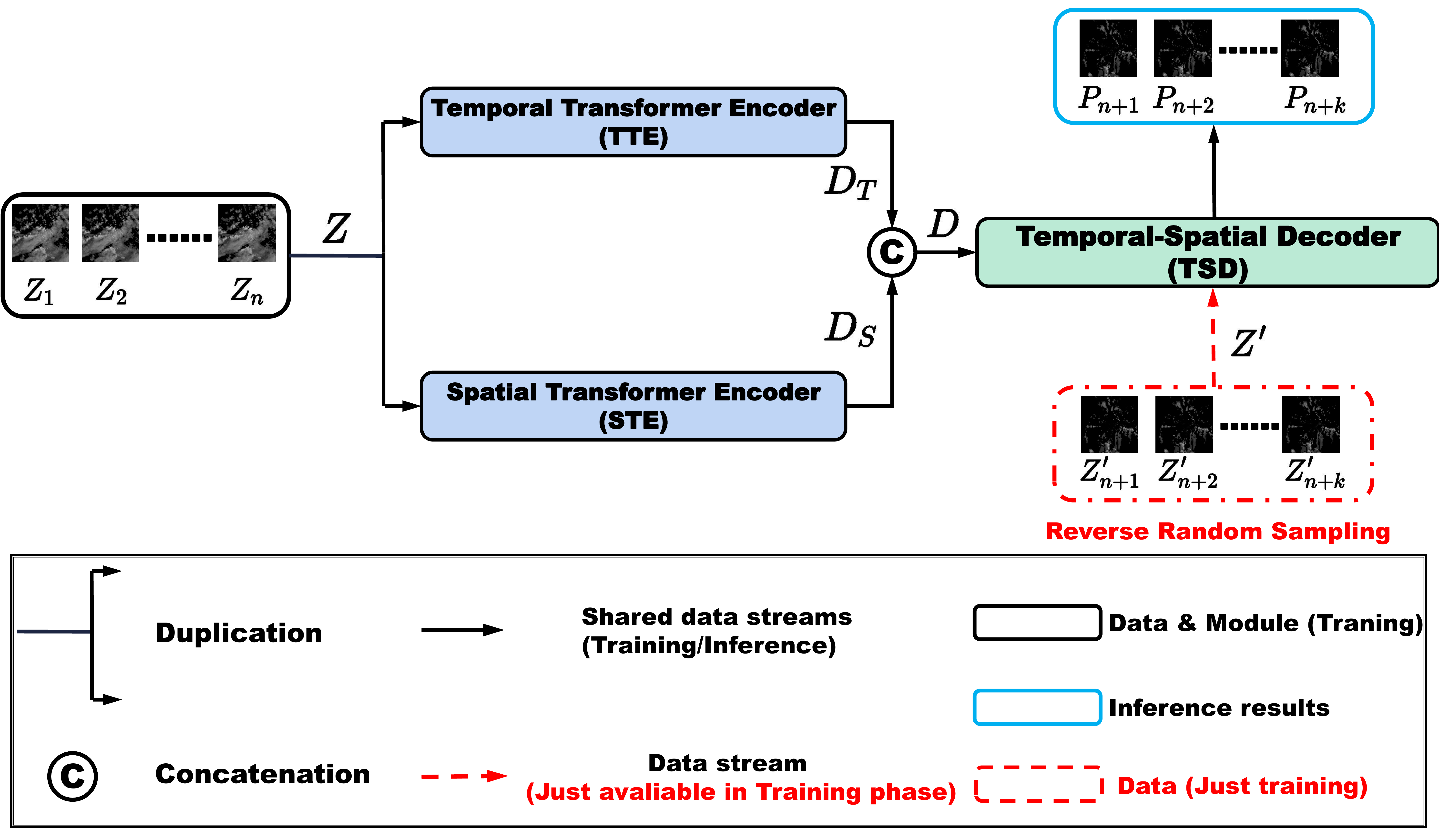}
\caption{Architecture of the proposed Temporal-Spatial Parallel Transformer, which consist of Temporal Transformer Encoder (TE), Spatial Transformer Encoder (SE), and Temporal-Spatial Decoder (TSD) in which the Reverse Random Sampling is utilized, where $Z_1, Z_2, \ldots, Z_n$ represent past historical observations, while $P_{n+1}, P_{n+2}, \ldots, P_{n+k}$ denote the extrapolation results for the future $k$ steps. $P'_{n+1}, P'_{n+2}, \ldots, P'_{n+k}$ serve as inputs for reverse random sampling, providing real prompts derived from the ground truth, which aims to encourage the model to generate more realistic images.}
\label{system}
\end{figure*}

The proposed Temporal-Spatial Parallel Transformer, known as \textit{TempEE}, is depicted in Fig. \ref{system}. Unlike auto-regression-based models that may encounter issues with cumulative error spreading during extrapolation, \textit{TempEE} addresses this concern by employing a one-step forward mechanism. The \textit{TempEE} comprises three main components, including Temporal Encoder (TE), Spatial Encoder (SE), and Temporal-Spatial Decoder (TSD). The TE and SE are responsible for extracting intricate spatio-temporal correlations of the echo features in the temporal and spatial dimensions, respectively. Subsequently, the TSD uses the concatenated spatio-temporal features to decode and predict future echoes. The forthcoming sections will provide a detailed description of each module.

\subsection{Temporal Encoder}
\begin{figure}[tbh]
\centering
\includegraphics[width=0.485\textwidth]{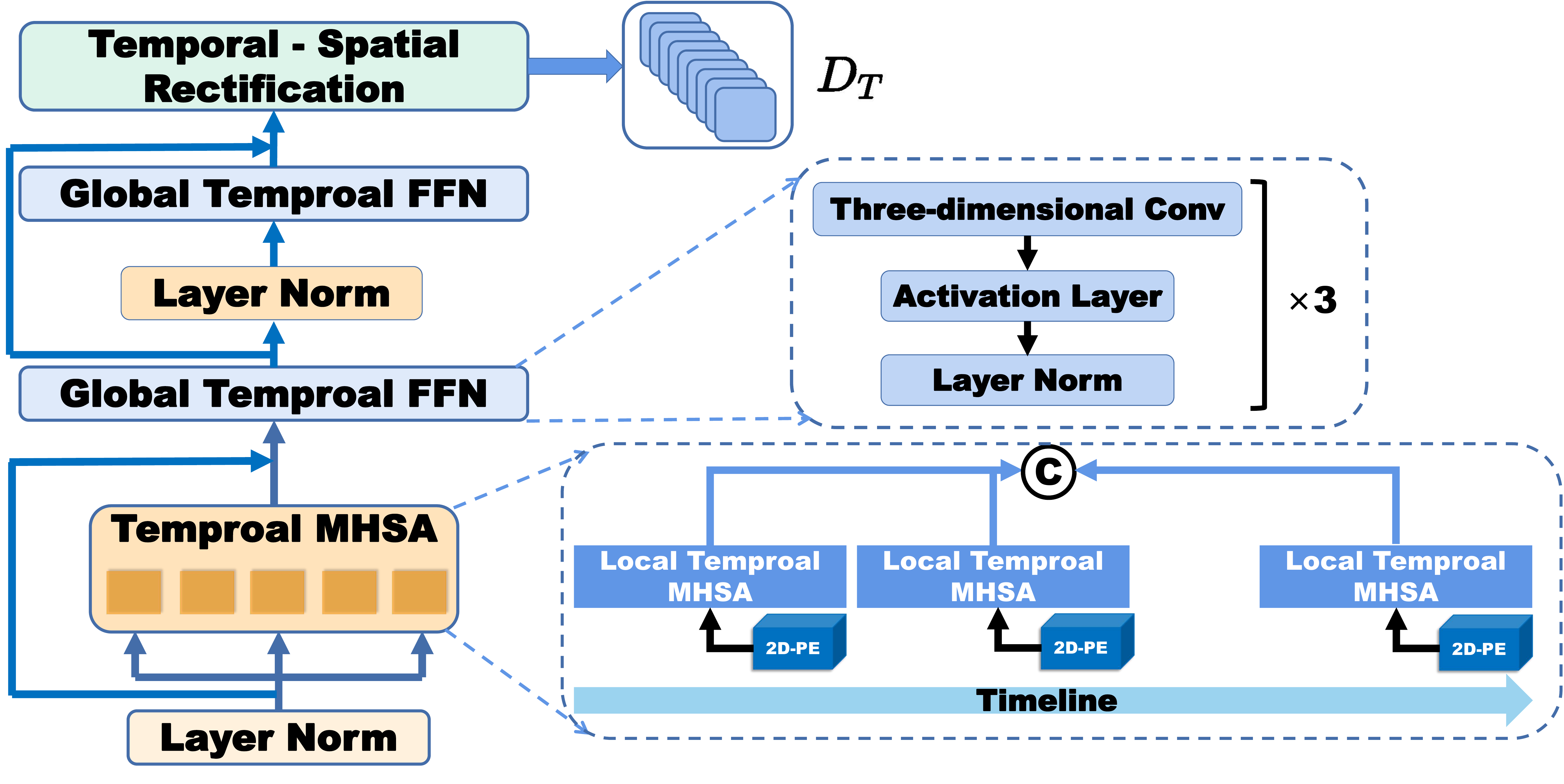}
\caption{Architecture of the Temporal Encoder (TE), which consists of two layer norm modules, two global Temporal FFN modules, Temporal Multi-Level Multi-Head Self-attention (Temporal MHSA), and Temporal-Spatial Rectification.}
\label{te}
\end{figure}

Our proposed Temporal Transformer Encoder (TE), is designed to accurately forecast echo images by effectively capturing their temporal correlations despite being low-resourced. As illustrated in Fig.~\ref{te}, the TE leverages a Temporal Multi-Level Multi-Head Self-Attention (Temporal MHSA) mechanism that employs weight-sharing and local Temporal MHSA modules to extract temporal correlation information from the echo feature along the timeline. The core of our TE model is the Temporal MHSA mechanism, which allows for uninterrupted learning and reshaping of local parameters to capture the degree of temporal correlation among the echo images. Our proposed TE model guarantees reliable forecasting performance while remaining resource-efficient.

\paragraph{MHSA in \textit{TempEE}} The success of Transformers is heavily reliant on its Multi-Head Self-Attention (MHSA), which is effective and efficient in capturing complex dependencies in input data. Nevertheless, it can introduce high computational complexity, resulting in slower inference speeds. This problem becomes particularly troublesome in real-time radar echo extrapolation-based weather forecasting. To address this challenge and ensure fast inference speed without compromising accuracy, we propose a novel approach called Multi-Level Multi-Head Self-Attention. This approach is designed to optimize the trade-off between complexity and accuracy.

\begin{figure}[tbh]
\centering
\includegraphics[width=0.4\textwidth]{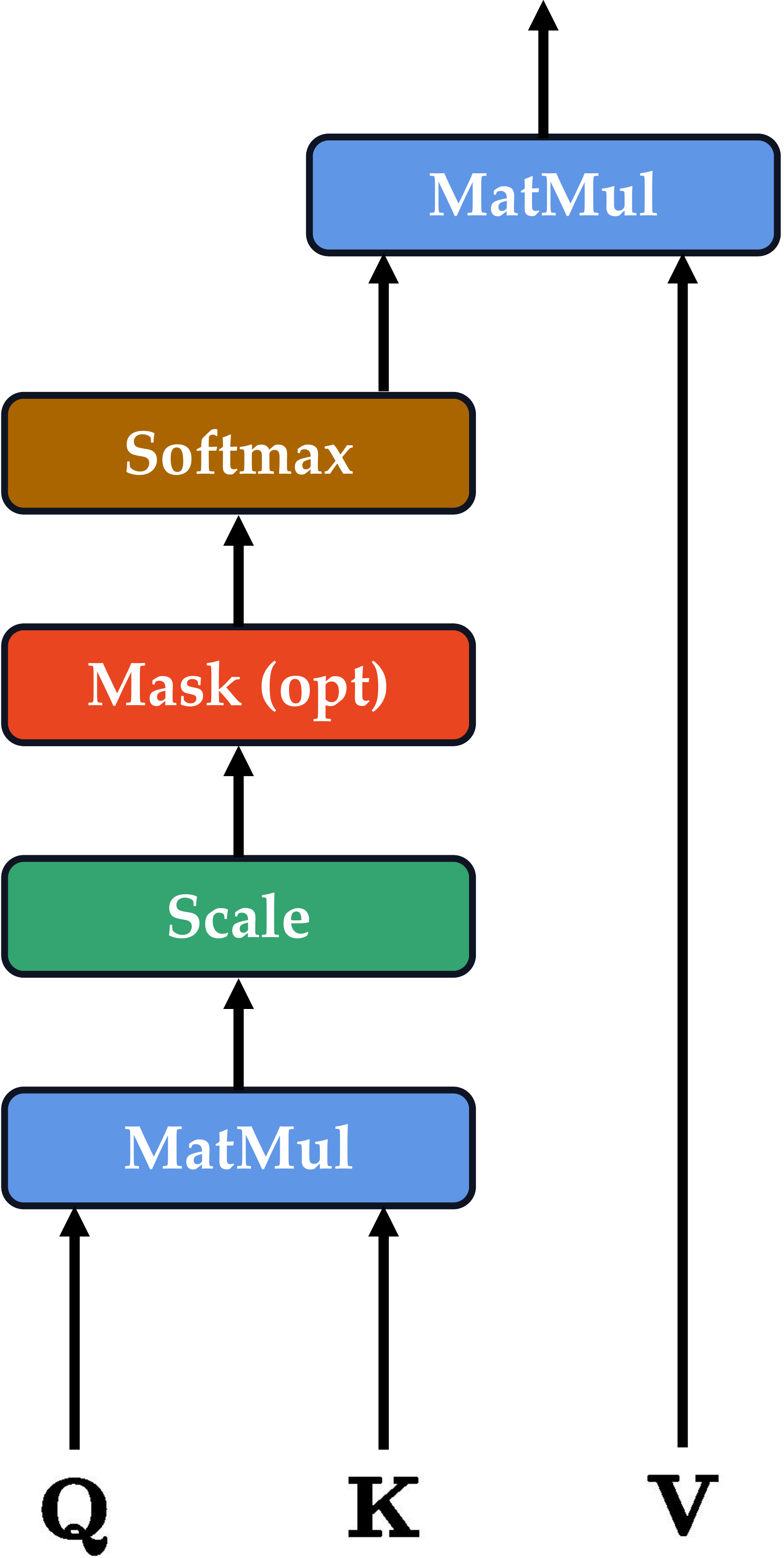}
\caption{Architecture of the proposed Multi-Level Multi-Head Self-Attention, where $\mathbf{Q}$ (Query), $\mathbf{K}$ (Key), and $\mathbf{V}$ (Value) are obtained by multiplying the inputs by each of the three different learnable parameter matrices. }
\label{cmhsa}
\end{figure}

The computation steps of the MHSA are introduced in the follows.
% Given the feature map $\mathbf{I_F} \in \mathbb{R}^{c \times h \times w}$, where $c$, $h$, and $w$ denotes the channels, height, and width, respectively. The $\mathbf{I_F}$ is reshaped before fed into the attention module to define and obtain the query $\mathbf{Q}$, key $\mathbf{K}$, and value $\mathbf{V}$, these can be formulated as follow: 
% \begin{equation}
% \mathbf{I_F} \in \mathbb{R}^{c \times h \times w} \xrightarrow{reshape} \mathbf{I_F} \in \mathbb{R}^{c \times (h \times w)},
% \label{eq1}
% \end{equation}
First, given the feature map $\mathbf{I_F} \in \mathbb{R}^{c \times h \times w}$, where $c$, $h$, and $w$ represent the channels, height, and width, respectively, $\mathbf{I_F}$ is reshaped before being fed into the attention module to obtain the query $\mathbf{Q}$, key $\mathbf{K}$, and value $\mathbf{V}$ matrices. Specifically, we reshape $\mathbf{I_F}$ to $\mathbf{I_F} \in \mathbb{R}^{c \times (h \times w)}$, and obtain:
\begin{equation}
\mathbf{Q}=W^q\mathbf{I_F}, \mathbf{K}=W^k\mathbf{I_F}, \mathbf{V}=W^v\mathbf{I_F},
\label{eq2}
\end{equation}
where the $W^q$, $W^k$, and $W^v$ denotes the learnable weight matrices for $\mathbf{Q}$, $\mathbf{K}$, and $\mathbf{V}$, respectively. Next, we compute the attention matrix from $\mathbf{Q}$ and $\mathbf{K}$ via Softmax:
\begin{equation}
\mathbf{Att}=Softmax(\mathbf{QK}^T/\sqrt{d}),
\label{eq3}
\end{equation}
where the $\sqrt{d}$ is an approximate normalization constant. The output is then computed by taking the matrix product of $\mathbf{Att}$ and $\mathbf{V}$, yielding $\mathbf{A}=\mathbf{Att}\mathbf{V}$. This process involves first computing the similarity between each pair of tokens using the product $\mathbf{QK}^T$, and then deriving each new token by combining all tokens based on their similarity. Finally, we compute the attention's output by adding a residual connection, which is expressed as
\begin{equation}
\begin{aligned}
\mathbf{I_F} \in \mathbb{R}^{c \times (h \times w)} &\xrightarrow{reshape} \mathbf{I_F} \in \mathbb{R}^{c \times h \times w},\\
&\mathbf{I_o} = \mathbf{A}\mathbf{W}^p,
\end{aligned}
\label{eq6}
\end{equation}
where the $\mathbf{W}^p$ is a trainable matrix that shape is $\mathbf{W}^p \in \mathbb{R}^{c \times c}$ for feature projection. 

Since this attention module contains $N$ heads internally (i.e., $N$-heads Self-Attention), its final attention output is summed over the channel dimensions according to the results of Eq. (9). The computation complexity of MHSA can be inferred from Eqs. (4)-(9), resulting in
\begin{equation}
\Omega(\text{MHSA})=2ch^2w^2+3hwc^2.
\label{mhsa}
\end{equation}
Additionally, the space complexity includes the term $O(h^2 w^2)$, indicating that memory consumption could become very large when the input is high-resolution ($h \times w$). This limits the application of the MHSA module to radar extrapolation tasks that place demands on fast training and inference.

\paragraph{Global Temporal FFN}
The Global Temporal FFN is a conventional network used for feature extrapolation that incorporates features from the preceding operation. This module comprises three sequences of layers: Layer Normalization, Activation (using \textit{ReLu}) and 3-D Convolution. A single part of Global Temporal FFN, denoted as \text{$GTF_i(\cdot)$} can be formulated as: $GTF_i(\cdot) = \text{Conv}(ReLu(\text{Layer Norm}(\cdot)))$.

\paragraph{Temporal-Spatial Rectification}
The proposed TE primarily focuses on establishing temporal correlations between echo images and the timeline. The Temporal-Spatial Rectification aims to rectify both the spatial and temporal echo features before feeding them into the encoder module with SE. This process combines the temporal and spatial features, enhancing the overall analysis. The transformation is represented as $\mathbf{I} \in \mathbb{R}^{c \times w \times h \times t} \rightarrow \mathbf{I} \in \mathbb{R}^{(c \times t) \times w}$, where $t$ denotes time steps.

\subsection{Spatial Encoder (SE)}
\begin{figure}[tbh]
\centering
\includegraphics[width=0.75\textwidth]{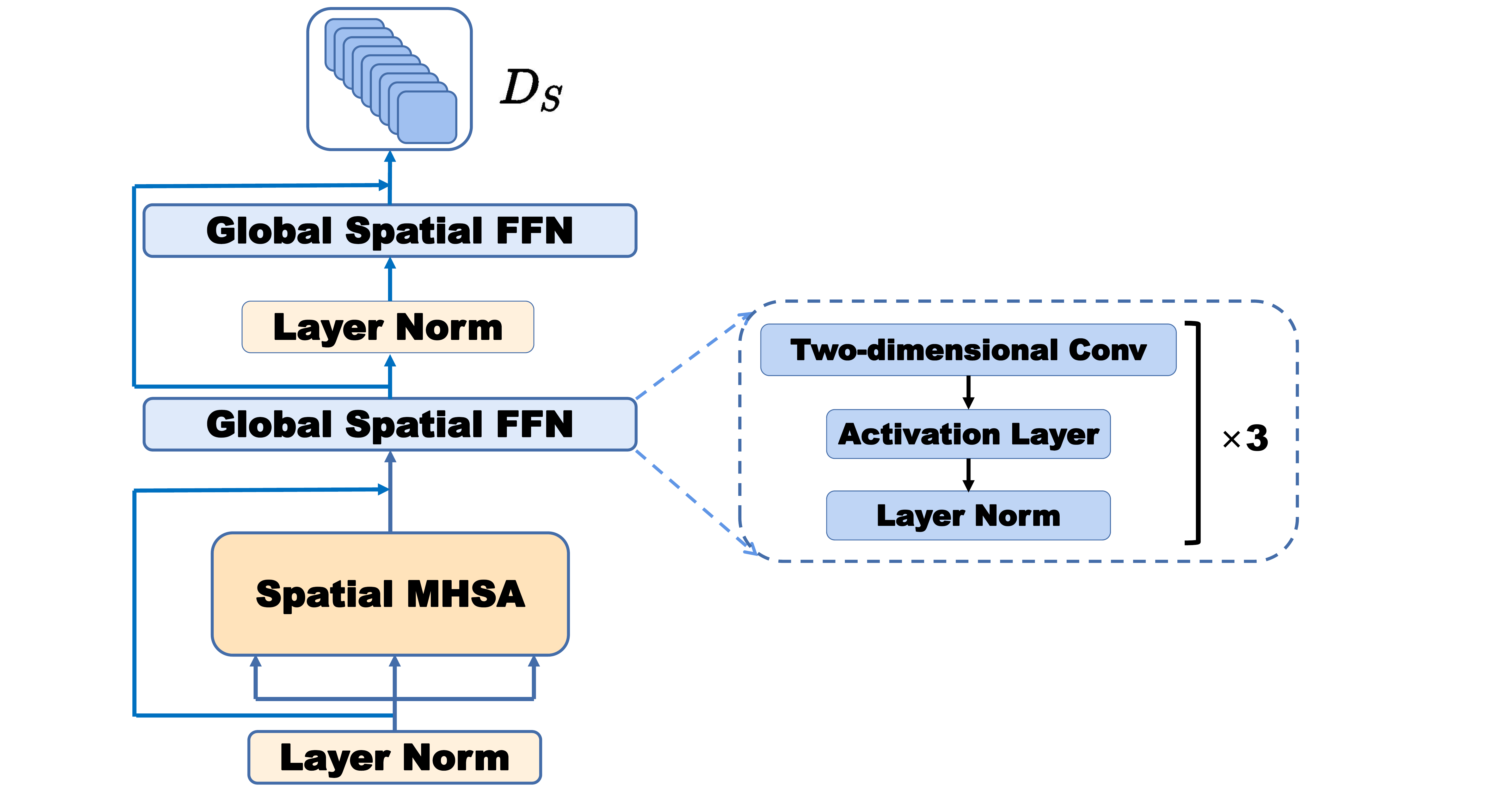}
\caption{Architecture of the Spatial Encoder (SE), which consists of two layer norm modules, two global Spatial FFN modules, and Spatial MHSA.}
\label{STE}
\end{figure}

The Spatial Encoder (SE) is designed to work in parallel with the Temporal Transformer Encoder (TE) to create contextual local-global feature correlations in echo images. Rather than attempting to challenge the understanding of forward temporal correlations along the timeline, the SE focuses on constructing these correlations. The structure of the SE is depicted in Fig.~\ref{STE}. Importantly, the Spatial MHSA eliminates the weight-sharing MHSA along the time-dimension, while the Temporal MHSA maintains consistency in all other aspects. Additionally, the convolutional module of Global Spatial FFN is replaced with 2-D Convolution units contrast to Global Temporal FFN. A single part of Global Spatial FFN ($GSF_i$) can be formulated as:
\begin{equation}
    GSF_i(\cdot) = \text{Conv}(ReLu(\text{Layer Norm}(\cdot))).
\end{equation}
\subsection{Temporal-Spatial Decoder (TSD)}
\begin{figure}[H]
\centering
\includegraphics[width=0.5\textwidth]{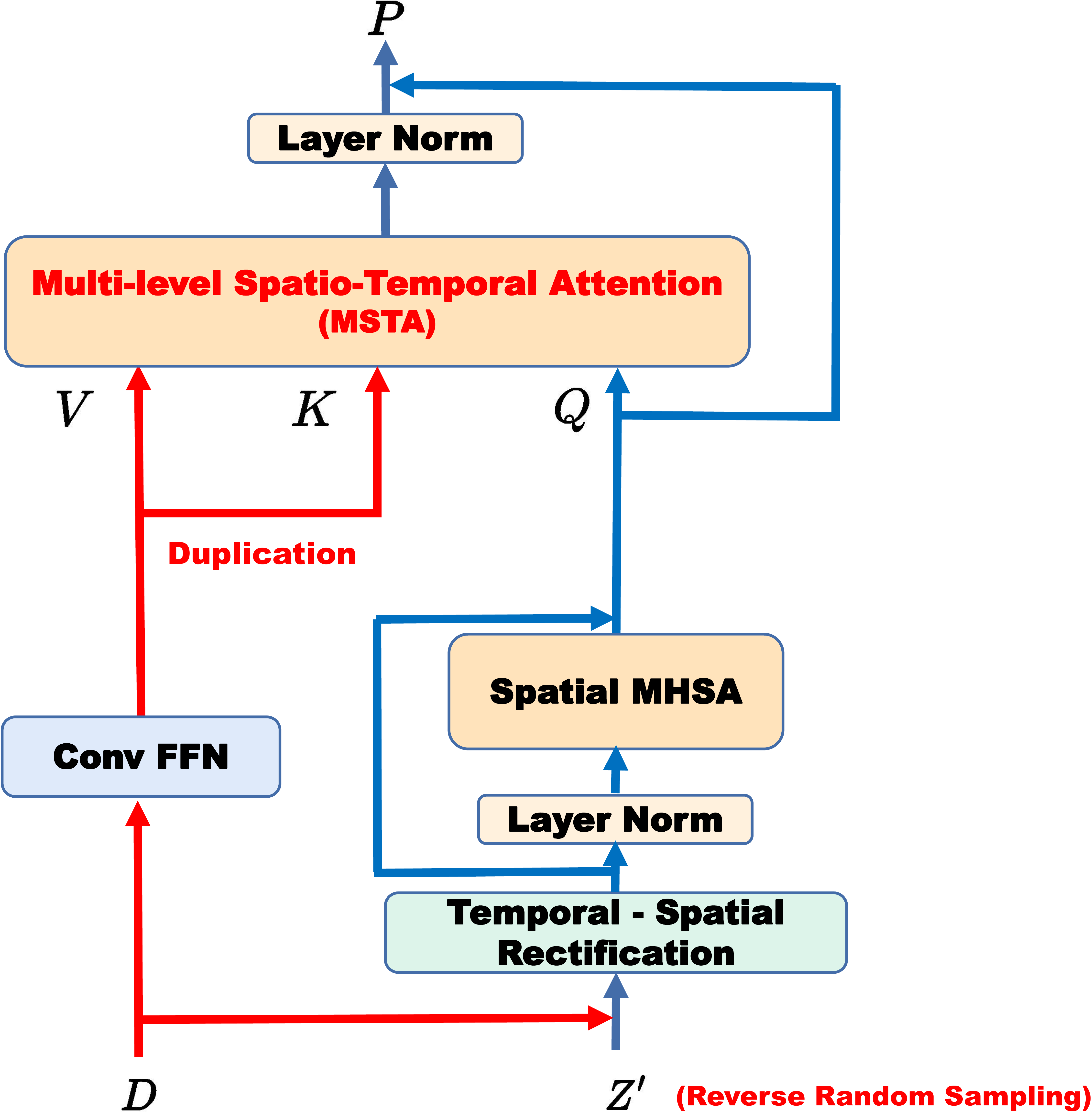}
\caption{Architecture of the Temporal-Spatial Decoder (TSD), which consists of a Conv FFN, a Temporal-Spatial Rectification, two Layer Norm modules, and Multi-level Spatio-Temporal Attention (MSTA), where $\mathbf{Q}$ (Query), $\mathbf{K}$ (Key), and $\mathbf{V}$ (Value) are obtained by multiplying the inputs by each of the three different learnable parameter matrices.}
\label{tsd}
\end{figure}

The Temporal-Spatial Decoder (TSD) marks a significant departure from the conventional Transformer Decoder's auto-regression. Instead, it employs a Multi-level Spatio-Temporal Attention (MSTA) and a reverse random sampling strategy to construct contextual relationships among sequences, enabling one-step forecasting. This approach tackles the problem of feature dissipation commonly observed in auto-regressive processes. The TSD's structure is illustrated in Figure \ref{tsd}. The subsequent section presents a theoretical analysis and design of the TSD, including Reverse Random Sampling and MSTA.

\paragraph{Reverse Random Sampling}
Additionally, the reverse random sampling strategy based on Ref.~\cite{wang2022predrnn} is introduced, which is used at the encoding timesteps and foreces the model to learn more about long-term dynamics by randomly hiding real observations with decreasing probabilities as the training phase progresses. We fix the last frame as $Z_{n+k}$, and the rest of the frames follow the sampling method as input to the TSD. This approach enables the model to be trained with auxiliary information/knowledge (the fixed last frame and the randomly sampled echo maps in the rest time period), which helps to capture the temporal dynamics between the echoes. During the testing phase, the sampling strategy becomes inapplicable to prevent any potential leakage of unseen data.

\begin{figure}[H]
\centering
\includegraphics[width=0.69\textwidth]{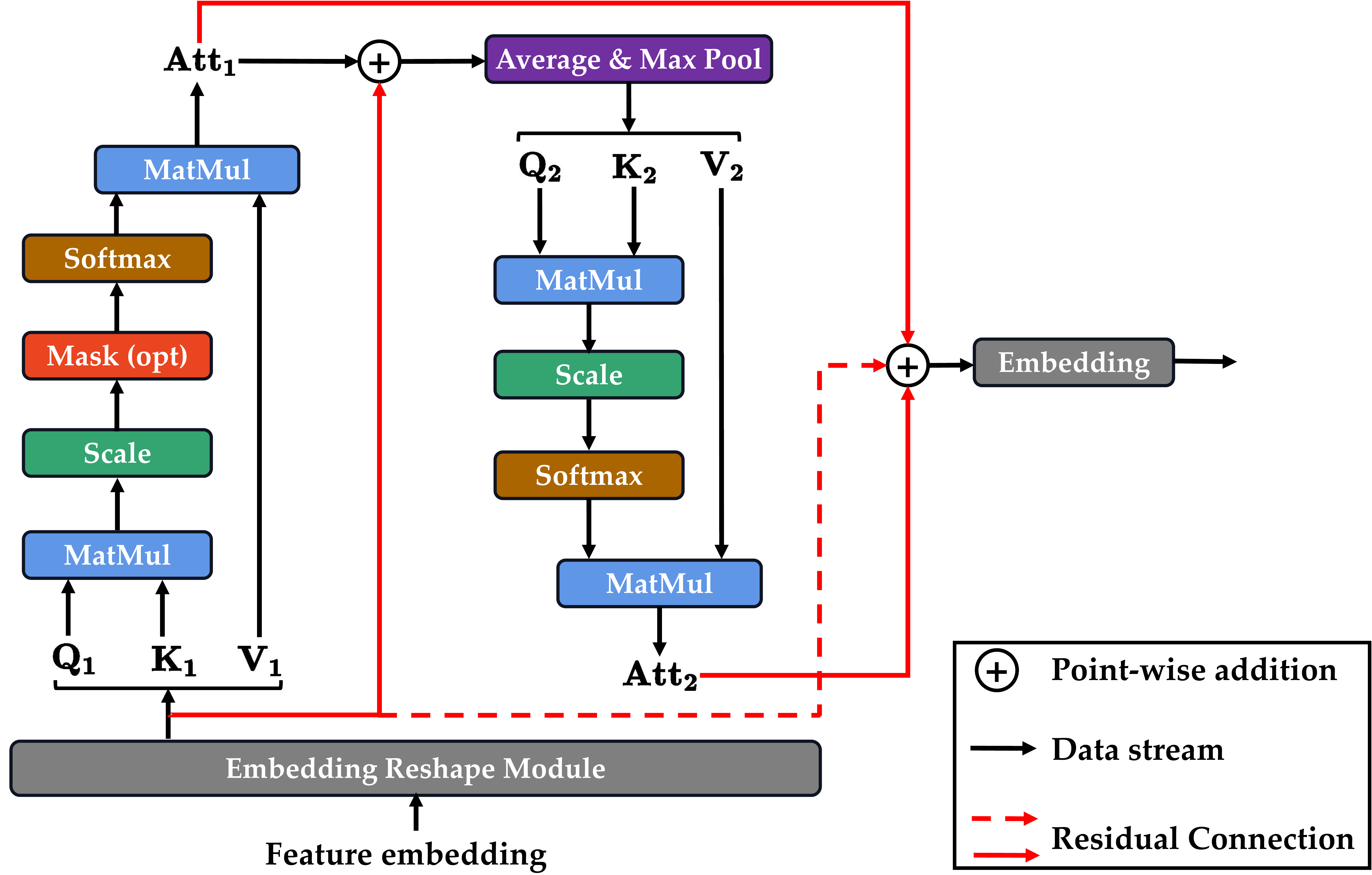}
\caption{Architecture of the proposed Multi-Level Multi-Head Self-Attention.}
\label{proMMHS}
\end{figure}

\paragraph{Multi-level Spatio-Temporal Attention (MSTA)}
Fig.\ref{proMMHS} depicts the architecture of the Multi-scale and multi-level Spatio-Temporal Attention (MSTA) method. Unlike the Multi-Head Self-Attention (MHSA) method shown in Fig.\ref{cmhsa}, which computes the attention score for the entire image simultaneously, MSTA processes the input echo image using a novel approach. MSTA divides the image into multiple progressively refined levels, with only a limited number of tokens involved in each computation. This approach allows MSTA to analyze fine-grained feature regions, such as weak intensity areas and typhoon eyes, while also capturing global-local information during the decoding process.

To achieve this, MSTA divides the input echo image $\mathbf{I_F} \in \mathbb{R}^{c \times h \times w}$ into smaller grids before computing the attention score. Each grid is represented as $\mathbf{G} \in \mathbb{R}^{g \times g}$, and the input feature map is reshaped accordingly. This strategy enables MSTA to concentrate on specific image regions instead of processing the whole image simultaneously.

The transformation process from $\mathbf{I_F}$ to $\mathbf{I_{F1}}$ is as follows: $\mathbf{I_F} \in \mathbb{R}^{c \times h \times w}$ is divided into $\mathbf{I_{F1}} \in \mathbb{R}^{c \times (\frac{h}{g} \times g) \times (\frac{w}{g} \times g)}$ small grids, which are further reshaped to $\mathbf{I_{F1}} \in \mathbb{R}^{c \times (\frac{h}{g} \times \frac{w}{g}) \times (g \times g) }$.
The query, key, and value of this level can be obtained by
\begin{equation}
\mathbf{Q_1}=\mathbf{I_{F1}}\mathbf{W}^{q}_{1}, \mathbf{K_1}=\mathbf{I_{F1}}\mathbf{W}^{k}_{1}, \mathbf{V_1}=\mathbf{I_{F1}}\mathbf{W}^{v}_{1},
\end{equation}
where $\mathbf{W}^{q}_{1}$, $\mathbf{W}^{k}_{1}$, and $\mathbf{W}^{v}_{1}$ are trainable weight matrices that is $\mathbf{W}^{q}_{1} \in \mathbb{R}^{c \times c}$, $\mathbf{W}^{k}_{1} \in \mathbb{R}^{c \times c}$, and $\mathbf{W}^{v}_{1} \in \mathbb{R}^{c \times c}$, respectively. Then, the local attention of this level can be generated by
\begin{equation}
    \mathbf{Att_1}=Softmax(\mathbf{Q_{1} K_{1}}^T/\sqrt{d})\mathbf{V_1},
\end{equation}

The input attention map $\mathbf{Att_1}$ is first reshaped to match the shape of the input image $\mathbf{I_F}$ before computing the subsequent level of local attention. A residual connection is then applied between this local attention map and the input image, resulting in the modified attention map $\mathbf{Att_1}$ given by:
\begin{equation}
\begin{aligned}
\mathbf{Att_1} \in \mathbb{R}^{c \times (\frac{h}{g} \times \frac{w}{g}) \times (g \times g)} &\rightarrow \mathbf{Att_1} \in \mathbb{R}^{c \times (\frac{h}{g} \times g) \times (\frac{w}{g} \times g)} \\
&\rightarrow \mathbf{Att_1} \in \mathbb{R}^{c \times h \times w },
\end{aligned}
\end{equation}
\begin{equation}
\mathbf{Att_1} = \mathbf{Att_1} + \mathbf{I_F}.
\end{equation}
Compared to the conventional MHSA, which operates directly on $h$ or $w$, $\mathbf{Att_1}$ performs attention on a smaller $g \times g$ grid, significantly reducing its space complexity. After entering the next computational level, $\mathbf{Att_1}$ is downsampled by pooling to $g'$. Each $g' \times g'$ small grid $\mathbf{G'} \in \mathbb{R}^{g' \times g'}$ is then treated as a token in the calculation. The process is expressed as:
\begin{equation}
\mathbf{Att_1'} = \frac{1}{2}[\alpha \times \text{MaxPool}{g'}(\mathbf{Att_1}) + \beta \times \text{AvePool}{g'}(\mathbf{Att_1})],
\end{equation}
where $\text{MaxPool}{g'}(x)$ and $\text{AvePool}{g'}(x)$ downsample the input feature map by $g'$ using max pooling and average pooling with filter size and stride of $g'$, respectively. The control coefficients $\alpha$ and $\beta$ are importance coefficients. The resulting $\mathbf{Att_1'}$ has shape $\mathbf{Att_1'} \in \mathbb{R}^{c \times \frac{h}{g'} \times \frac{w}{g'}}$. The token size $(\frac{h}{g'} \times \frac{w}{g'})$ can be obtained from the reshaped $\mathbf{Att_1'}$:
\begin{equation}
\begin{aligned}
\mathbf{Att_1'} \in \mathbf{R}^{c \times \frac{h}{g'} \times \frac{w}{g'}} \rightarrow \mathbf{Att_1'} \in \mathbf{R}^{c \times (\frac{h}{g'} \times \frac{w}{g'})}.
\end{aligned}
\end{equation}

At this level, $\mathbf{Att_1'}$ computes the query, key, and value, which are denoted as $\mathbf{Q}_2$, $\mathbf{K}_2$, and $\mathbf{V}_2$, respectively:
\begin{equation}
\mathbf{Q_2}=\mathbf{Att_1'}\mathbf{W}^{q}_{2}, \quad \mathbf{K_2}=\mathbf{Att_1'}\mathbf{W}^{k}_{2}, \quad \mathbf{V_2}=\mathbf{Att_1'}\mathbf{W}^{v}_{2},
\end{equation}
where $\mathbf{W}^{q}_{2}$, $\mathbf{W}^{k}_{2}$, and $\mathbf{W}^{v}_{2}$ are trainable weight matrices with dimensions $\mathbf{W}^{q}_{2} \in \mathbb{R}^{c \times c}$, $\mathbf{W}^{k}_{2} \in \mathbb{R}^{c \times c}$, and $\mathbf{W}^{v}_{2} \in \mathbb{R}^{c \times c}$, respectively. The local attention $\mathbf{Att_2}$ can be computed as follows:
\begin{equation}
\mathbf{Att_2}=\mathrm{Softmax}(\mathbf{Q_{2} K_{2}}^T/\sqrt{d})\mathbf{V_{2}},
\end{equation}
The final output of the MSTA is computed by the reshaping operation and a residual connection:
\begin{equation}
\mathbf{Att_2} \in \mathbb{R}^{c \times (h \times w)} \rightarrow \mathbf{Att_2} \in \mathbb{R}^{c \times h \times w},
\end{equation}
\begin{equation}
\mathrm{MSTA}(\mathbf{I_F}) = (\mathbf{Att_1} + \mathbf{Att_2})\mathbf{W}^m \mathbf{W}^n + \mathbf{I_F},
\end{equation}
where $\mathbf{W}^{m}$ and $\mathbf{W}^{n}$ are trainable weight matrices. By fusing different levels of attention outputs, the proposed MMSA has strong local and global feature modeling capabilities. The computational complexity of MMSA can be computed as follows:
\begin{equation}
\Omega(\mathrm{MSTA})=chw(2g'^2 + 4c) + \frac{2chw}{g'^2}(c+hw),
\label{mmsa}
\end{equation}
which implies that the computational complexity is reduced from $O(h^2 w^2)$ to $O(hwg^2 + \frac{h^2 w^2}{g'^2})$.
The computation complexity of Multi-Head Self-Attention can be expressed:
\begin{equation}
\Omega(\text{MHSA})=2ch^2w^2+3hwc^2.
\end{equation}
According to the Eq. (13), the comparison of computational complexity between our proposed MSTA and the conventional MHSA is cleared. Suppose that the echo image input is $360\times360\times3$, and $g'$ in MSTA is 2, the computational parameters of conventional MHSA and our MSTA are:
\begin{equation}
\begin{aligned}
  \text{P}_{\text{MHSA}} = 2*3*360^4+3*360^2*3^2 = 100,780,459,200,\\
  \text{P}^{g'=2}_{\text{MSTA}} = 3*360^2*(2^3+4*3) + \frac{2*3*360^2}{2^2}*(3+360^2) \\ = 25,202,599,200,\\
\end{aligned}
\end{equation}
This means that our MSTA pays only a quarter of the computational cost of the conventional MHSA when $g'=2$.

\section{Experiment}
\label{sec_exp}
\subsection{Dataset and Pre-processing}
\begin{figure}[H]
\centering
\includegraphics[width=0.75\textwidth]{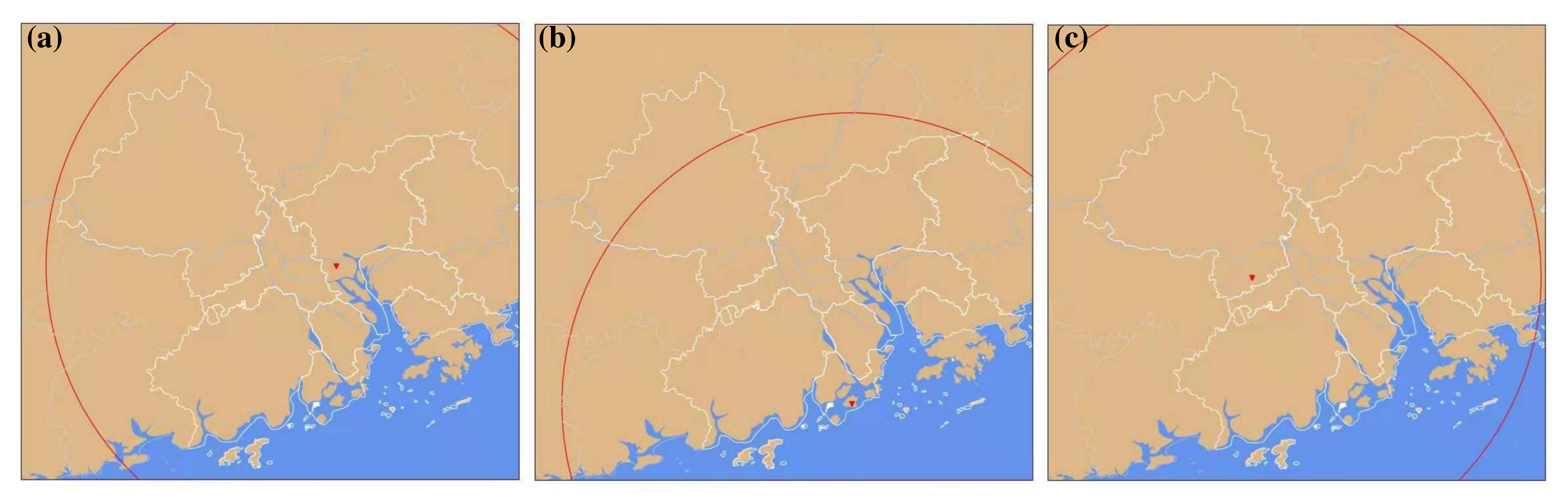}
\caption{Location and radiation range of radar: (a) Guangzhou; (b) Zhaoqing; (c) Zhuhai. The lower red triangle represents the position of the radar, and the red circle represents the radiation range.} 
\label{F9} 
\end{figure}

% \paragraph{Foshan Dataset}
Radar echo images were collected from three independent radars which were located in Guangzhou, Zhaoqing, and Zhuhai, Guangdong Province, China. All radar echoes are centered in Foshan, Guangdong Province, China, and the radiation area is $360 \times 360$km, as shown in Fig. \ref{F9}. These radars echoes are Constant Altitude Plan Position Indicator (CAPPI) that are collected at six-minute intervals at 1,000 meters, and data spans from January 5, 2018, to May 28, 2021.

\begin{figure}[H]
\centering
\includegraphics[width=0.75\textwidth]{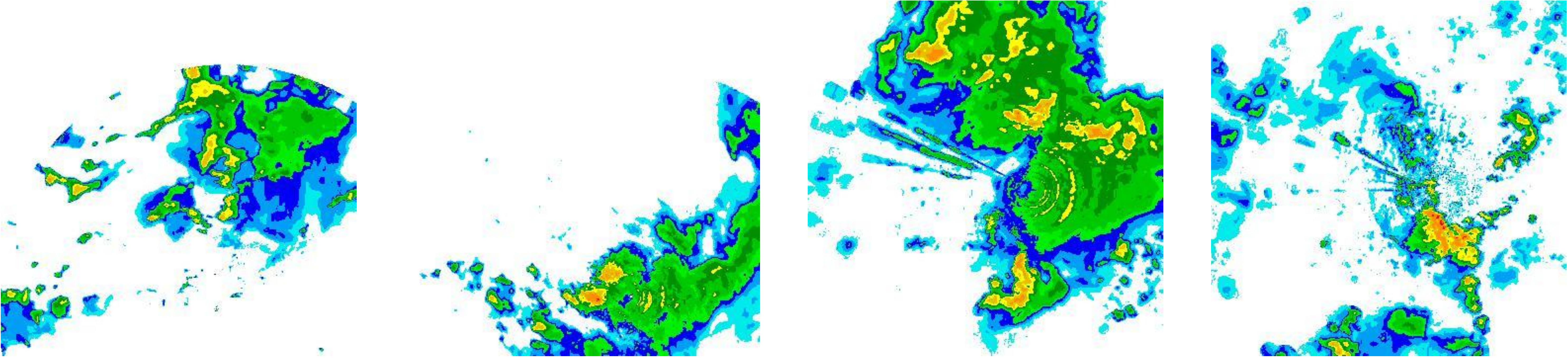}
\caption{Radar echo image examples of Foshan Dataset.} 
\label{imageexamples} 
\end{figure}

Fusion of the echo images of the three radars is a vital initial data preprocessing step, ensuring the coherence between the areas covered by the echo distribution in the network input and output. This step is mathematically expressed as:
\begin{equation}
    R_{(x, y)}=Max[GZ_{(x, y)},ZQ_{(x, y)}, ZH_{(x, y)}]
\end{equation}
where $R_{(x, y)}$, $GZ_{(x, y)}$, $ZQ_{(x, y)}$ and $ZH_{(x, y)}$ represent the intensity of the final dataset radar echo, Guangzhou radar echo, Zhaoqing radar echo, and Zhuhai radar echo at point $(x, y)$ respectively. $x, y$ are considered integers were belonging to the interval $[0, 360]$ because the size of the radar echo image is 360 × 360. Note that they will be resized to 224×224 during the training. Some examples are shown in Fig.~\ref{imageexamples}.
After pre-processing, the radar echo images are divided into three distinct sets. The first set comprises radar echo images captured between 2018 and 2020, which serves as the training and validation dataset, following a ratio of 9:1 for proper training. The second set consists of radar echo images solely from the year 2021, reserved exclusively for the test dataset. Importantly, there is no data overlap among the training, validation, and test sets.

% \paragraph{GuangZhou Dataset}
% Some radar echo image of Guangzhou Dataset are shown in Figure. All echo image's size in this dataset is $480 \times 480$, which means that each image cover the area of $480 km \times 480 km$.

% \begin{figure}[H]
% \centering
% \includegraphics[width=0.48\textwidth]{Example2.pdf}
% \caption{Radar echo image examples of Guangzhou Dataset.} 
% \label{xas} 
% \end{figure}
\subsection{Baseline}
We introduce five baseline models for traditional and DL-based radar extrapolation to demonstrate the effectiveness of our proposed \textit{TempEE}: Optical-flow, MIM, PredRNN, PredRNN++, and CMS.

\textbf{Optical-flow}: a approach used to track the apparent motion of objects in consecutive frames, which estimates the displacement vectors of pixels between frames, enabling the analysis of object movement and scene dynamics~\cite{woo2017operational}.

\textbf{MIM}: a new approach for spatiotemporal prediction using Memory In Memory (MIM) networks, which utilize differential signals to model non-stationary and approximately stationary properties~\cite{wang2019memory}.

\textbf{PredRNN}: a recurrent network that models visual dynamics with compositional subsystems and achieves highly competitive results on five datasets for spatiotemporal sequence predictive learning.~\cite{wang2017predrnn,wang2022predrnn}.

\textbf{PredRNN++}: an improved version of the PredRNN model, which includes two additional components: a spatial attention mechanism and a progressive growing mechanism~\cite{wang2018predrnn++}.

\textbf{CMS}: a novel approach to spatiotemporal predictive learning that focuses on context correlations and multi-scale spatiotemporal flow using two elaborately designed blocks: Context Embedding and Spatiotemporal Expression~\cite{chai2022cms}.

To ensure a fair and unbiased comparison, both the baseline models and \textit{TempEE} undergo training in an identical environment. This includes utilizing the same dataset and maintaining consistent experimental settings. Additionally, we keep the network structure and parameter settings of all baseline models same with those specified in the corresponding original publications to ensure uniformity and comparability.

\subsection{Experimental Setup}
All radar echo were resized to $192 \times 192$, and the batch size is set to 12 in experiments. To reduce computational consumption without losing features, the original images were chunked using the patch operation with a patch value of 12. This paper aims to utilize a historical dataset comprising 20 frames to predict the corresponding 20 future frames. During training, we employed the \textit{AdamW} optimizer along with the WarmUp mechanism to avoid setting an excessively high initial learning rate, which could result in unstable training. The WarmUp mechanism gradually increases the learning rate from a smaller value to a preset maximum value at the beginning of training. This stepwise adjustment ensures a more stable and effective learning process, as
\begin{equation}
    lr_t = \frac{k^{d_{max}}}{\sqrt{d_t}}\times min(\frac{t}{w},\sqrt{t})
\end{equation}
where $lr_t$ denotes the learning rate of the $t$th step, $d_t$ denotes the hidden layer dimension of the $t$th step, $d_{max}$ denotes the maximum dimension, $k$ denotes a constant, and $w$ denotes the number of WarmUp steps.

To enhance the training effectiveness, a learning rate Cosine decay operation is utilized at the end of the WarmUp. This ensures that the learning rate reduces stationaryly throughout the training process and prevents the problem of the rate decreasing too fast, causing unstable training. Moreover, Cosine decay also permits the model to maintain a smaller learning rate in later stages of training, facilitating better exploration of the parameter space and mitigating overfitting. The expression of this operation is as follows:
\begin{equation}
    lr_t = \frac{1}{2}(1 + \cos(\frac{T_{cur}}{T_{max}}\pi))\times lr_{max},
\end{equation}
where $lr_t$ denotes the learning rate of the $t$-th step, $T_{cur}$ denotes the the current step, $T_{max}$ denotes the total number of steps, and $lr_{max}$ denotes the maximum learning rate. Additionally, the Early-Stopping strategy is also employed during training to mitigate the risk of overfitting. The patience is set to 10, which means the training will be terminated once the loss on the validation set does not decrease for ten consecutive iterations.

\subsection{Performance Evaluation Metric}
\paragraph{Image-based Metric}
The Mean Squared Error (MSE) measures the average squared difference between the original image and the reconstructed image. A lower MSE value indicates better image quality, it can be expressed as:
\begin{equation}
    \mathcal{L} = \frac{1}{Q\cdot mn}\sum_{T=0}^{Q}\sum_{i=0}^{m}\sum_{j=0}^{n}[I_{i, j}-K_{i, j}]^2,
\end{equation}
where the $I$ is ground truth of echo image and $K$ is the extrapolation, $Q$ denotes the extrapolation period, $m$ and $n$ is the echo image's row and column, respectively.

Peak Signal-to-Noise Ratio (PSNR) is a commonly used measure in image reconstruction research. It represents the ratio between the maximum possible power of a signal and the power of corrupting noise that affects the fidelity of the signal's representation. PSNR is expressed in decibels (dB), and a larger value indicates better image quality.
\begin{equation}
    \text{PNSR} = 10\log_{10}\left(\frac{MAX_I^2}{\text{MSE}}\right),
\end{equation}
where $MAX_I$ is $255$ in our experiments.

The Structural Similarity Index (SSIM) measures the similarity between two images by comparing their luminance, contrast, and structural information. SSIM compares the local image structures between the original and the reconstructed images, rather than just comparing the pixel values. A higher SSIM value indicates better image quality.
\begin{equation}
    \text{SSIM}(x,y) = \frac{(2\mu_x\mu_y+c_1)(2\sigma_{xy}+c_2)}{(\mu_x^2+\mu_y^2+c_1)(\sigma_x^2+\sigma_y^2+c_2)},
\end{equation}
where $\mu_x$ and $\mu_y$ are the means of $x$ and $y$, respectively, $\sigma_x^2$ and $\sigma_y^2$ are the variances of $x$ and $y$, respectively, $\sigma_{xy}$ is the covariance of $x$ and $y$, and $c_1$ and $c_2$ are two constants.

Learned Perceptual Image Patch Similarity (LPIPS) metric employs learned features extracted by neural networks to quantify the perceptual similarity between a pair of images. % The basis of LPIPS lies in the hierarchical processing of visual information in the human visual system, which begins with lower-level features like edges and textures and advances to higher-level features linked to object recognition.
LPIPS calculates the distance between the feature representations of two images, with their features learned from a dataset of images. A lower LPIPS value indicates a higher degree of perceptual similarity between the images.
\begin{equation}
    \text{LPIPS}(x,y) = \sum_{i=1}^{N}w_i\cdot\|\phi_i(x)-\phi_i(y)\|_2,
\end{equation}
where $\phi_i(\cdot)$ denotes the feature map output from the $i$th layer of the model, $w_i$ denotes the weight of the $i$th layer feature map, and $N$ denotes the number of layers of the model.

\paragraph{Meteorological Evaluation Metric}
We utilized commonly accepted evaluation metrics in meteorology to synthetically assess model performance, such as the probability of detection (POD), false alarm rate (FAR), and critical success index (CSI). These metrics provide a measure of the model's prediction accuracy. A lower FAR value indicates a larger discrepancy between the extrapolated echo and the ground truth. Higher values of CSI, POD, and accuracy (ACC) demonstrate a closer correlation between the predicted results and ground truth. Under perfect forecast conditions, the FAR would be zero, and the CSI, POD, and ACC would all be equal to one.

We calculated these metrics based on rainfall intensities, ranging from 0.05 mm/h to 40 mm/h, estimated from the radar echo reflectance using the Z-R relationship~\cite{anagnostou1999real}, and all radar echo map are converted into images by:
\begin{equation}
    V_{pixel} = [255\times \frac{V_{dBZ}}{70} + 0.5]
\end{equation}
where the $V_{pixel}$ and $V_{dBZ}$ denote the pixel value of echo image and intensity of echo maps, respectively. The threshold determines whether a certain pixel value on the echo image becomes 0/1 based on the echo intensity in the area represented by the point. The parameters True Positive (TP), Fake Negative (FN), Fake Positive (FP), and True Negative (TN) can be calculated by comparing predicted results to ground truth. The meteorological evaluation metrics were formulated as:
\begin{equation}
\begin{aligned}
    \text{POD} &=\frac{\text{TP}}{\text{TP}+\text{FN}},\\
    \text{FAR} &= \frac{\text{TP}}{\text{TP}+\text{FP}},\\
    \text{CSI} &=\frac{\text{TP}}{\text{TP}+\text{FN}+\text{FP}},\\
    \text{ETS} &= \frac{\text{TP}}{\text{TP} + \text{FN} + \text{FP} - \text{TN}}.
\end{aligned}
\end{equation}

\subsection{Experiment Results}
To verify the effectiveness and superiority of the proposed \textit{TempEE}, a series of experiments were conducted and they were analyzed quantitatively and qualitatively.

\paragraph{Image-based Performance Evaluation} Radar echo extrapolation involves using historical observations to predict future radar representations. In this section, we evaluate the proposed \textit{TempEE}'s effectiveness in generating high-quality radar images compared to baseline models. To assess the validity and superiority of \textit{TempEE}, we utilized image-based metrics, including PSNR, SSIM, MSE, and LPIPS, to perform a comprehensive evaluation. The performance comparison between \textit{TempEE} and the baseline is presented in Table I. The results demonstrate that \textit{TempEE} significantly outperforms the other baseline models, indicating that our proposed approach is capable of achieving superior accuracy and image quality for temporal-spatial image generation in radar echo extrapolation.
\begin{table}[tbh]
  \centering
      \caption{Performance comparison of the proposed \textit{\textit{TempEE}} and baseline under metrics on temporal-spatial image generation, and $\uparrow$ and $\downarrow$ represent the higher/lower the metric, the better/poorer the performance, the \textbf{Bold} means the optimal.}
    \begin{tabular}{ccccc}
    \toprule
    Model & PNSR $\uparrow$ & SSIM $\uparrow$ & MSE $\downarrow$ & LPIPS $\downarrow$ \\
    \midrule
    \midrule
    Optical-Flow & 16.70 & 0.450 & 539.63 & / \\
    PredRNN & 23.47 & 0.498 & 309.17 & 5.270 \\
    PredRNN++ & 22.90 & 0.515 & 301.93 & 4.188 \\
    CMS & 23.15 & 0.509 & 282.82 & 4.102 \\
    MIM & 23.11 & 0.520 & 249.16 & 4.232 \\
    \textit{\textit{TempEE}} (Ours) & \textbf{32.41} & \textbf{0.876} & \textbf{29.82} & \textbf{1.058} \\
    \bottomrule
    \end{tabular}%
  \label{tab:addlabel}%
\end{table}%

\paragraph{Quantitative Performance Evaluation of Perception Forecasting}
The performance of the proposed \textit{TempEE} model was evaluated under varying rainfall conditions using four meteorological Evaluation Metric: CSI, FAR, FAR, and POD. The model's performance was compared with a baseline model using the complete test dataset, and the results were analyzed and presented in Table II.

The analysis shows that the \textit{TempEE} model consistently outperforms the baseline model across different rainfall intensities. Specifically, as rainfall intensity increases, the \textit{TempEE} model exhibits the lowest decay span for each metrics, indicating a near unit-by-unit decrease in \textit{TempEE} of about 0.13 on CSI for rainfall intensity intervals ranging from 10 mm/h to 30 mm/h. In comparison, the corresponding figures for other models, such as Optical-flow, PredRNN, PredRNN++, CMS, and MIM, are 0.15, 0.16, 0.20, 0.18, and 0.18, respectively. These results indicate that our \textit{TempEE} model outperforms the baseline models in the overall extrapolation process. This means that the proposed \textit{TempEE} is highly effective in predicting rainfall under different conditions, as demonstrated by its superior performance in comparison to the baseline models across all rainfall intensities.

\begin{figure}[H]
\centering
\includegraphics[width=0.95\textwidth]{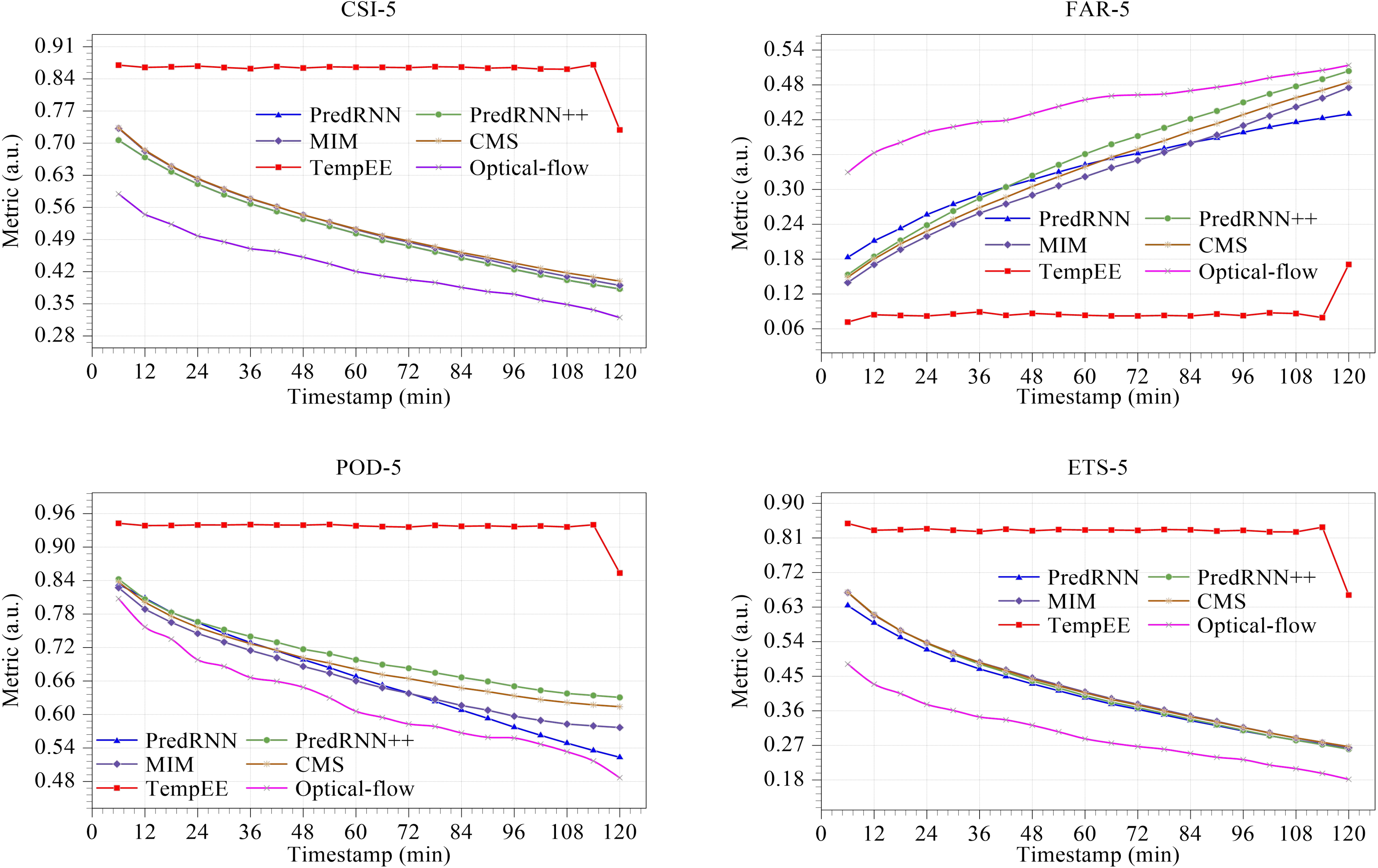}
\caption{Time-by-time performance comparison of the proposed \textit{TempEE} and baseline model under precipitation forecasting-based metrics when $\tau = 5$.} 
\label{Rainfall5} 
\end{figure}

\begin{figure}[tbh]
\centering
\includegraphics[width=0.95\textwidth]{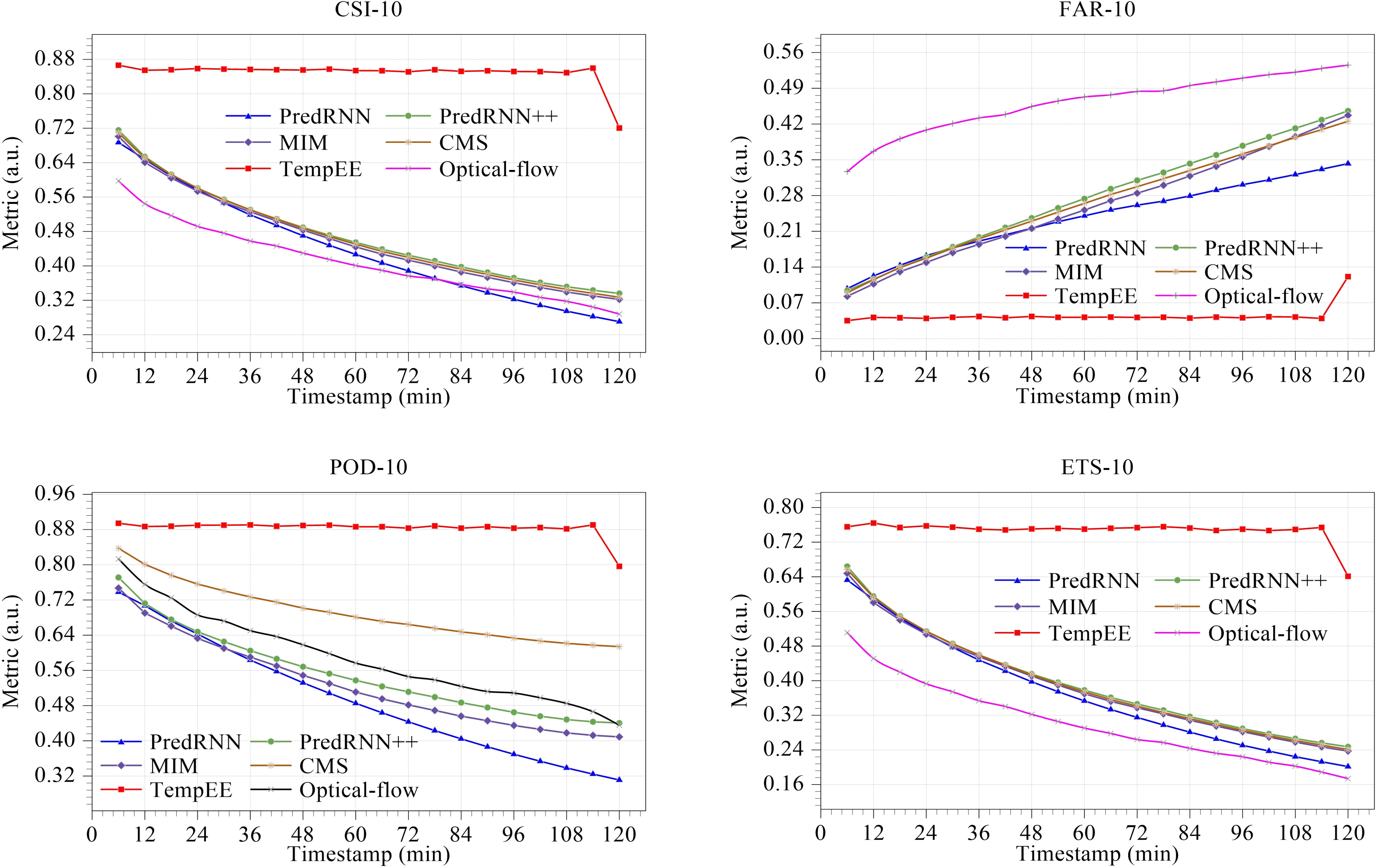}
\caption{Time-by-time performance comparison of the proposed \textit{TempEE} and baseline model under precipitation forecasting-based metrics when $\tau = 10$.} 
\label{Rainfall10} 
\end{figure}

\begin{figure}[tbh]
\centering
\includegraphics[width=0.95\textwidth]{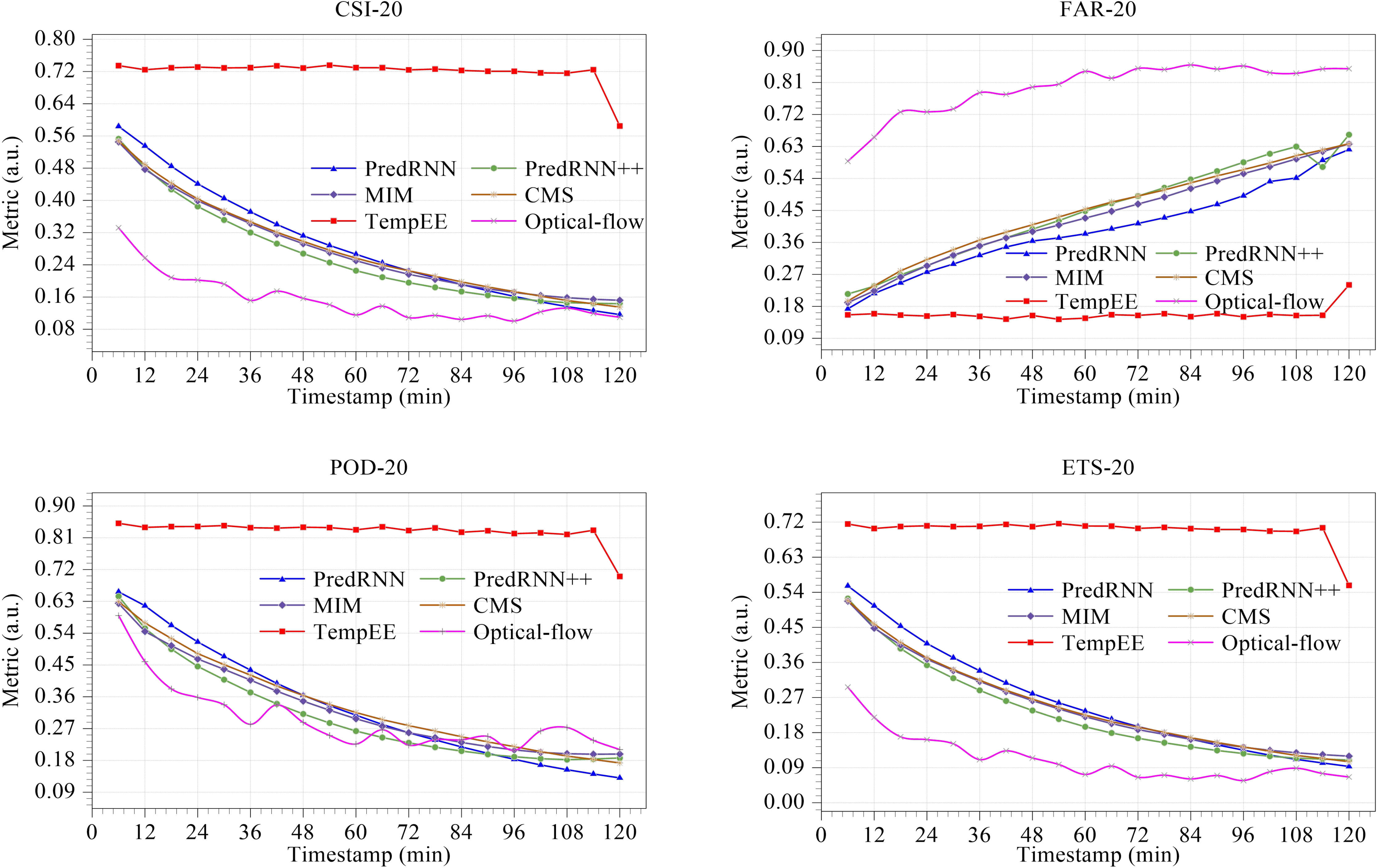}
\caption{Time-by-time performance comparison of the proposed \textit{TempEE} and baseline model under precipitation forecasting-based metrics when $\tau = 20$.} 
\label{Rainfall20} 
\end{figure}

\begin{figure}[tbh]
\centering
\includegraphics[width=0.95\textwidth]{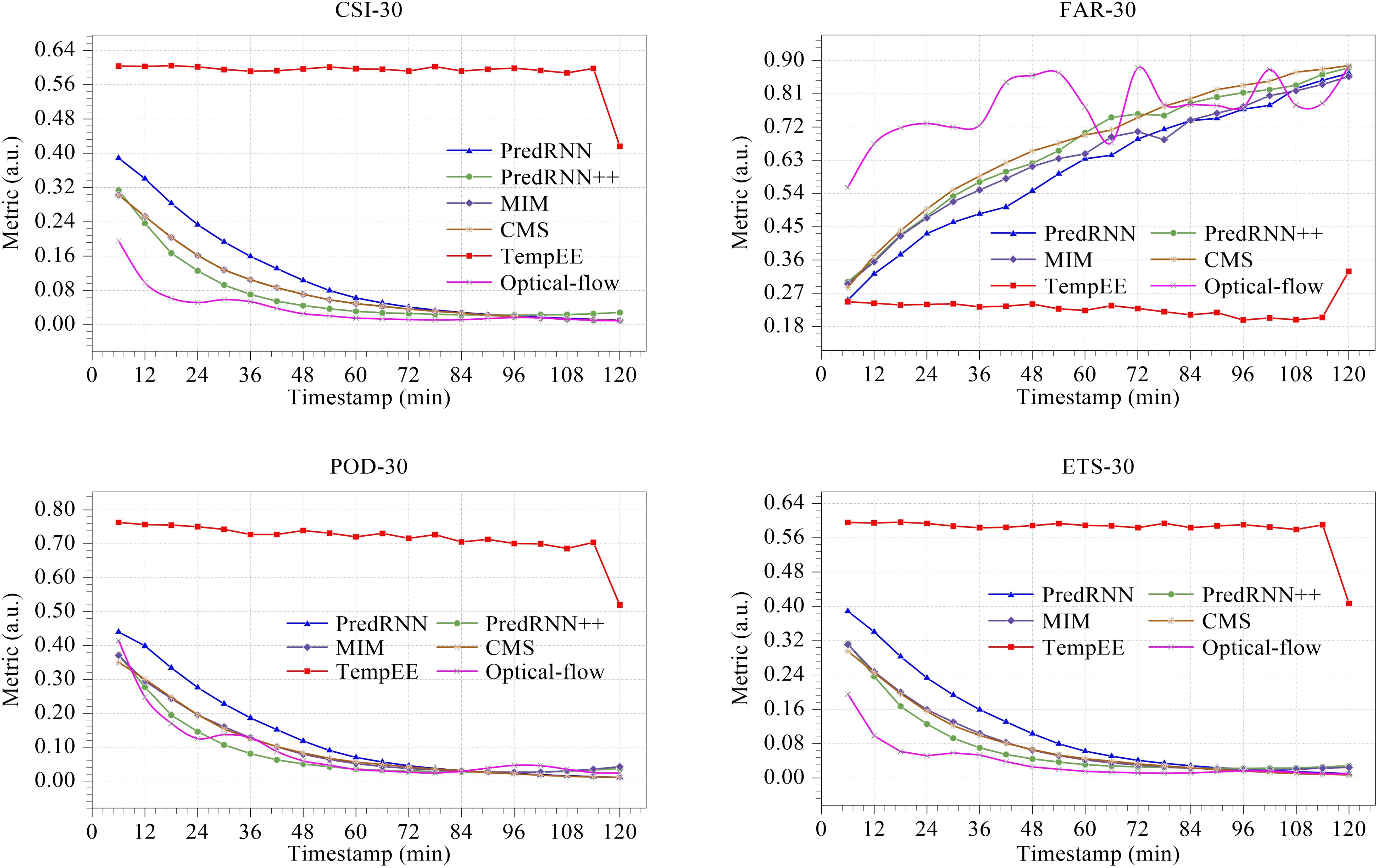}
\caption{Time-by-time performance comparison of the proposed \textit{TempEE} and baseline model under precipitation forecasting-based metrics when $\tau = 30$.} 
\label{Rainfall30} 
\end{figure}

\begin{figure}[H]
\centering
\includegraphics[width=0.95\textwidth]{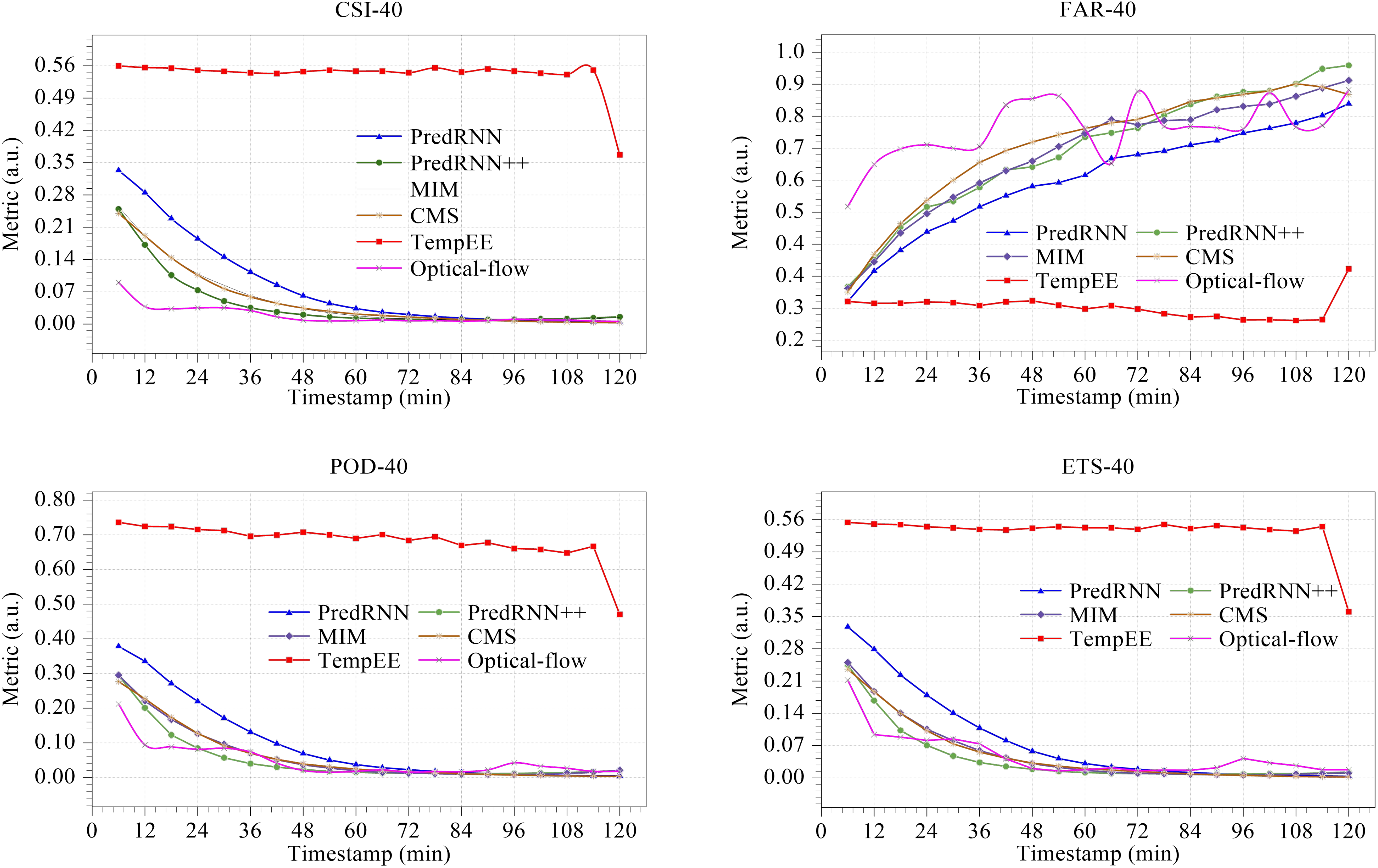}
\caption{Time-by-time performance comparison of the proposed \textit{TempEE} and baseline model under precipitation forecasting-based metrics when $\tau = 40$.} 
\label{Rainfall40} 
\end{figure}

One potential drawback of radar extrapolation algorithms is cumulative error spreading, which can significantly degrade the accuracy and quality of echo in long-term extrapolation. To assess the effectiveness of our \textit{TempEE} in long-term extrapolation, we compare its performance to the baseline over time at rainfall intensities of 5 mm/h, 10 mm/h, 20 mm/h, 30 mm/h, and 40 mm/h in Fig.\ref{Rainfall5}, Fig.\ref{Rainfall10}, Fig.\ref{Rainfall20}, Fig.\ref{Rainfall30}, and Fig.\ref{Rainfall40}, respectively. Our analysis indicates that: (1) \textit{TempEE} outperforms baseline models across different evaluation metrics and rainfall intensities;
(2) \textit{TempEE} exhibits the smoothest changes in metrics. In contrast, Optical-flow shows highly fluctuating FAR curves when predicting 30 and 40 mm rainfall, attributed to its weak perception of the echo kinematics; (3) \textit{TempEE} can maintain relatively stationary performance in long extrapolation processes compared to other models that suffer from the decay of accuracy over time. The only exceptions are the large fluctuations observed at the first time node ($0 \sim 6$ min) and the last time node ($114 \sim 120$ min), which are caused by the out random reverse sampling strategy that mixes prior knowledge at the first time node and the last point of time node. Another reason is that the fixed last-frame echo maps, which serve as auxiliary information, are not applicable during the testing process. However, this does not compromise the excellent extrapolation quality and generation performance of the model. This indicates that \textit{TempEE}'s one-step forward extrapolation is superior to traditional auto-regression and avoids error spreading during the extrapolation.

\begin{table}[H]
  \centering
  \caption{Performance comparison between the proposed \textit{TempEE} and baseline based on perception forecasting-based metrics, where $\tau$ is the evaluation threshold representing different rainfall intensities, and $\uparrow$ and $\downarrow$ represent the higher/lower the metric, the better/poorer the performance, the \textbf{Bold} means the optimal.}
  \resizebox{1\textwidth}{!}{
    \begin{tabular}{ccccccccccccr}
    \toprule
    Model/Metrics & \multicolumn{6}{c}{CSI $\uparrow$}     & \multicolumn{6}{c}{ETS $\uparrow$} \\
    \midrule
    Threshold   & $\tau = 5$ & $\tau = 10$ & $\tau = 20$ & $\tau = 30$ & $\tau = 40$ & \multicolumn{1}{c}{Average}  & $\tau = 5$ & $\tau = 10$ & $\tau = 20$ & $\tau = 30$ & $\tau = 40$ & \multicolumn{1}{c}{Average} \\
    \midrule
    Optical-flow & 0.410 & 0.385 & 0.156 & 0.078 & 0.030 & 0.212 & 0.281 & 0.276 & 0.131 & 0.031 & 0.012 &  0.146\\
    PredRNN & 0.511 & 0.438 & 0.288 & 0.111 & 0.082 & 0.286 & 0.397 & 0.373 & 0.254 & 0.106 & 0.078 &  0.242\\
    PredRNN++ & 0.522 & 0.469 & 0.261 & 0.071 & 0.044 & 0.273 & 0.413 & 0.398 & 0.259 & 0.067 & 0.061 &  0.240\\
    CMS & 0.524 & 0.466 & 0.282 & 0.084 & 0.051 & 0.271 & 0.411 & 0.391 & 0.272 & 0.119 & 0.049 &  0.248 \\
    MIM & 0.521 & 0.45 & 0.276 & 0.086 & 0.054 &  0.277 & 0.419 & 0.400 & 0.246 & 0.077 & 0.069 & 0.242 \\
    \textit{TempEE} (Ours) & \textbf{0.861} & \textbf{0.849} & \textbf{0.719} & \textbf{0.588} & \textbf{0.540} &  \textbf{0.705}  & \textbf{0.822} & \textbf{0.746} & \textbf{0.699} & \textbf{0.579} & \textbf{0.534} &  \textbf{0.676} \\
    \midrule
    Model/Metrics & \multicolumn{6}{c}{FAR $\downarrow$}   & \multicolumn{6}{c}{POD $\uparrow$} \\
    \midrule
    Threshold        & $\tau = 5$ & $\tau = 10$ & $\tau = 20$ & $\tau = 30$ & $\tau = 40$ &
    \multicolumn{1}{c}{Average}  & $\tau = 5$ & $\tau = 10$ & $\tau = 20$ & $\tau = 30$ & $\tau = 40$ & \multicolumn{1}{c}{Average} \\
    \midrule
    Optical-flow & 0.443 & 0.468 & 0.795 & 0.812 & 0.895 & 0.683 & 0.568 & 0.540 & 0.216 & 0.065 & 0.041 &  0.286\\
    PredRNN & 0.237 & 0.337 & 0.476 & 0.615  & 0.754  &  0.484  & 0.672 & 0.493 & 0.307 & 0.128 & 0.094 & 0.338 \\
    PredRNN++ & 0.279 & 0.329 & 0.464 &  0.599  &  0.734  &  0.481   & 0.681 & 0.551 & 0.332 & 0.083 & 0.051 & 0.340 \\
    CMS & 0.268 & 0.337 & 0.449 &  0.643 & 0.772  &  0.494  & 0.691 & 0.541 & 0.339 & 0.108 & 0.05 & 0.346 \\
    MIM & 0.261 & 0.323 & 0.433 & 0.638 & 0.707 &  0.472  & 0.709 & 0.527 & 0.328 & 0.101 & 0.062 & 0.345 \\
    \textit{TempEE} (Ours) & \textbf{0.045} & \textbf{0.088} & \textbf{0.158} & \textbf{0.231} & \textbf{0.273} &  \textbf{0.159}  & \textbf{0.934} & \textbf{0.882} & \textbf{0.828} & \textbf{0.716} & \textbf{0.682} & \textbf{0.808} \\
    \bottomrule
    \end{tabular}}
  \label{tab:addlabel}%
\end{table}%

\begin{figure}[tbh]
\centering
\includegraphics[width=0.95\textwidth]{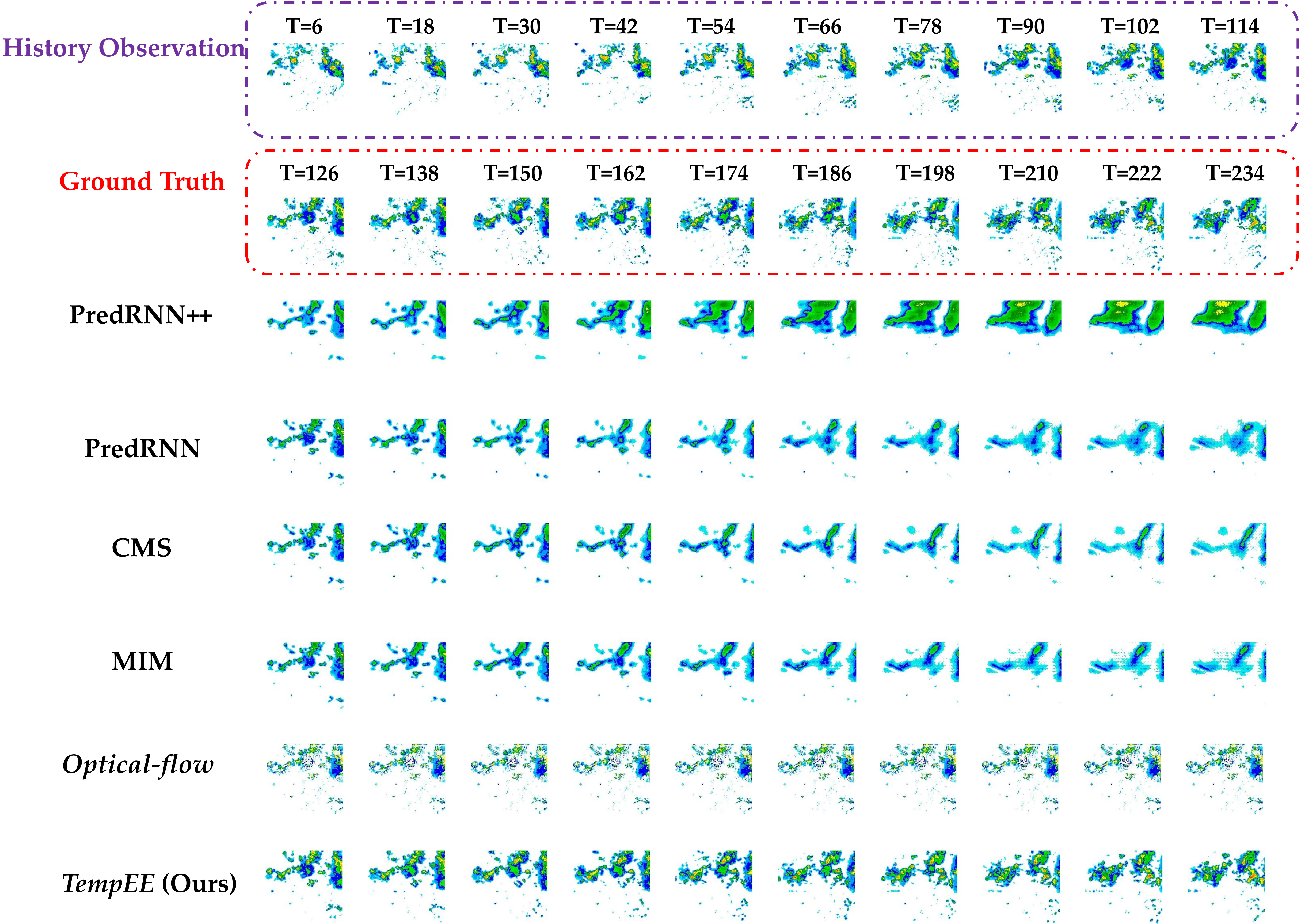}
\caption{Visualization of the proposed \textit{TempEE} and baseline models during the echo Sparse-Stationary motion inference process, note that the time period actually is 6 $min$, but 12 $min$ is used as the presentation period.} 
\label{Sparsestationary} 
\end{figure}

\begin{figure}[tbh]
\centering
\includegraphics[width=0.95\textwidth]{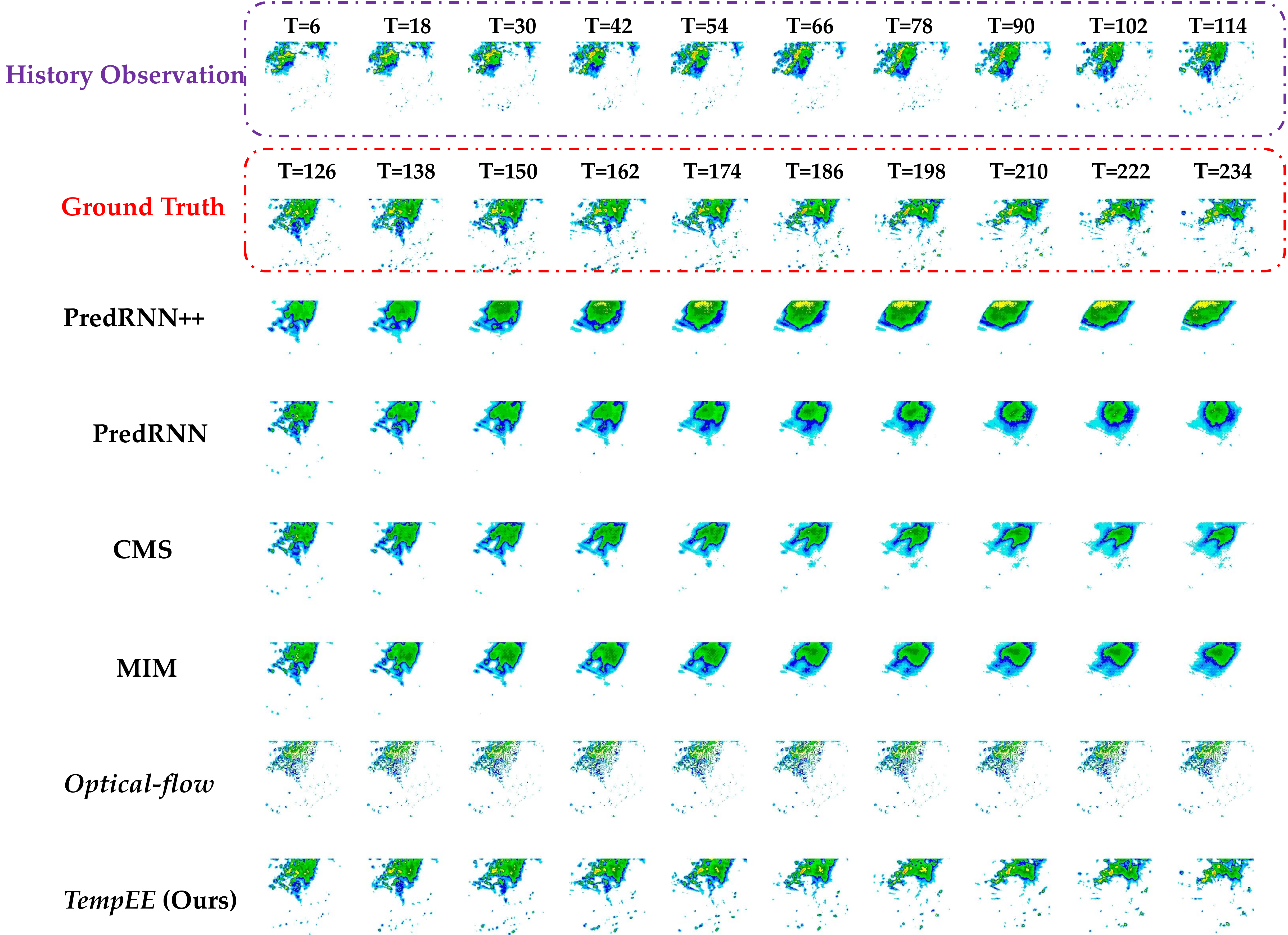}
\caption{Visualization of the proposed \textit{TempEE} and baseline models during the echo Dense-Stationary motion inference process, note that the time period actually is 6 $min$, but 12 $min$ is used as the presentation period.} 
\label{Densestationary} 
\end{figure}

\begin{figure}[tbh]
\centering
\includegraphics[width=0.95\textwidth]{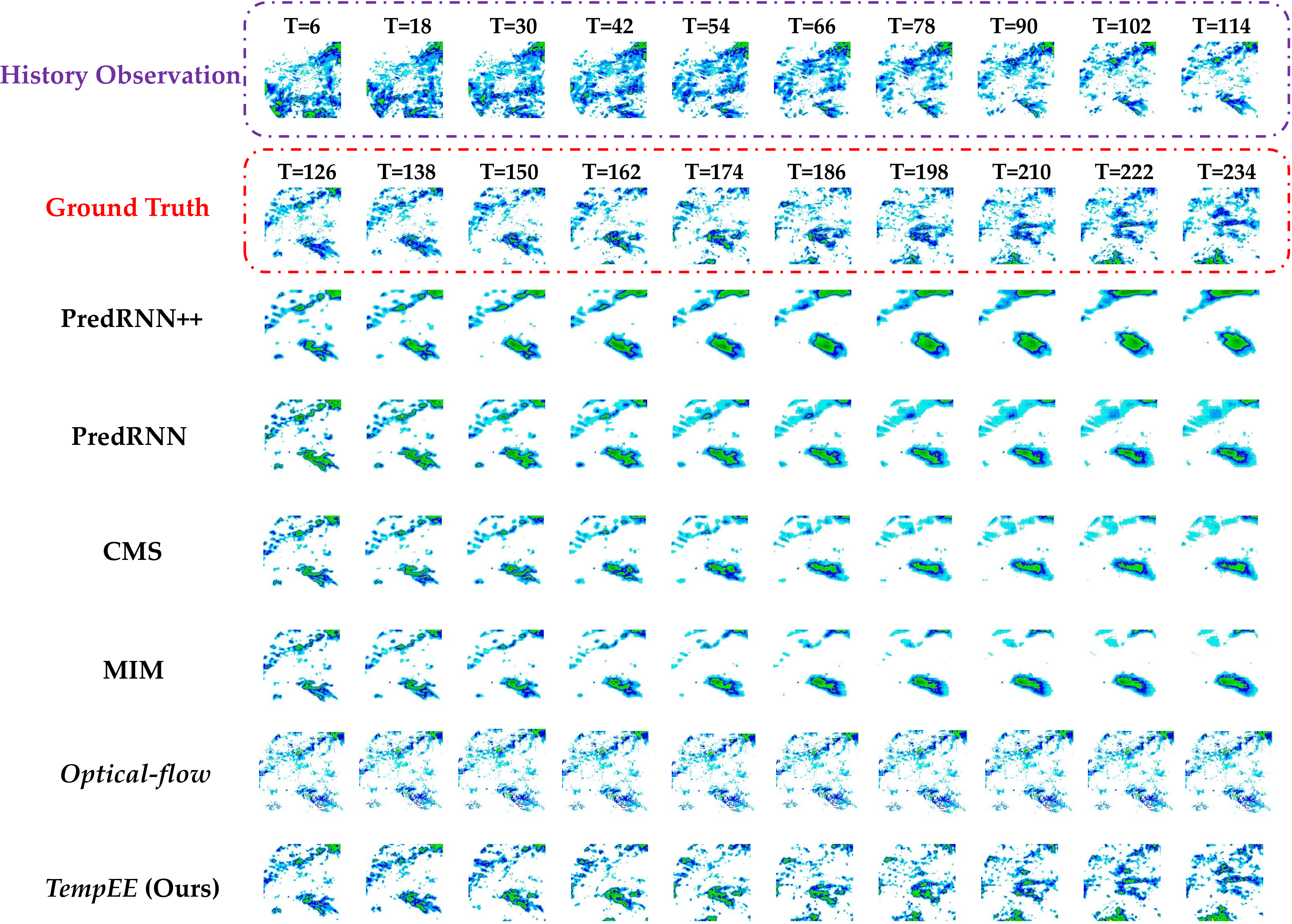}
\caption{Visualization of the proposed \textit{TempEE} and baseline models during the echo Sparse-NonStationary motion inference process, note that the time period actually is 6 $min$, but 12 $min$ is used as the presentation period.} 
\label{Sparse non-stationary} 
\end{figure}

\begin{figure}[tbh]
\centering
\includegraphics[width=0.95\textwidth]{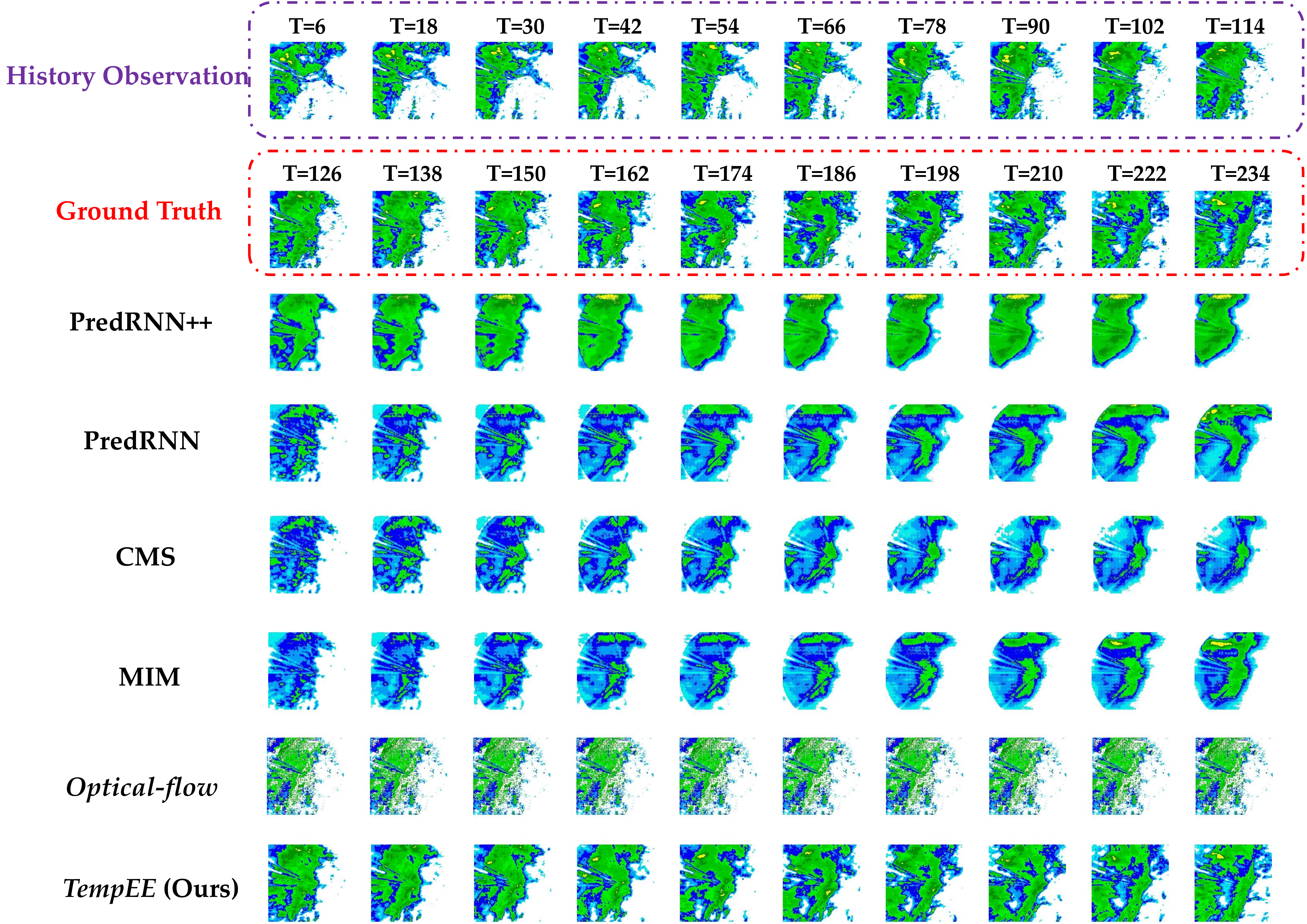}
\caption{Visualization of the proposed \textit{TempEE} and baseline models during the echo Dense-Nonstationary motion inference process, note that the time period actually is 6 $min$, but 12 $min$ is used as the presentation period.} 
\label{Dense non-stationary} 
\end{figure}

\paragraph{Qualitative Performance Evaluation of Perception Forecasting}
The quality of radar echo images is vital for precipitation forecasters to obtain effective guidance from extrapolation models. The effectiveness and prediction accuracy of the proposed \textit{TempEE} model were tested using four different echo motion processes from the complete test dataset. To distinguish between these processes, we categorized them into two groups: stationary and nonstationary. Moreover, based on the characteristics of the echoes, we further classified the processes into four subgroups: Sparse-Stationary (Fig.\ref{Sparsestationary}), Dense-Stationary (Fig.\ref{Densestationary}), Sparse-Nonstationary (Fig.\ref{Sparse non-stationary}), and Dense-Nonstationary (Fig.\ref{Dense non-stationary}). The visualization of these extrapolation results shouwcased the effectiveness and superior prediction accuracy of the proposed \textit{TempEE} model across various echo motion processes.

\noindent \textbf{Case Study on Spare-Stationary Motion Process.} During a two-hour sparse-stationary motion, the baseline models captured the general echo distribution, but they were unable to account for most of the sparse echo features, especially the series of sparse points in the lower right of the echo image. Over time, these baseline models demonstrated unsatisfactory results in both echo intensity and feature distribution. On the other hand, the proposed \textit{TempEE} model can effectively capture the fine echo features. Although there might be a few points with extrapolated intensity that are either too high or too low, its overall effect remains stable over time.

\noindent \textbf{Case Study on Dense-Stationary Motion Process.}
All models show the ability to capture the general motion trend of the radar echo distribution. For example, from 126 to 234 $min$, the echo gradually moves from the left to the upper right direction, accompanied by a certain degree of radar echo dissipation. The Baseline model captures the overall motion trend, but it has difficulty accounting for the finer details present in the data. Moreover, as the extrapolation process advances, the Baseline model fails to capture the radar echo feature dissipation.

In contrast, the proposed \textit{TempEE} model offers a more comprehensive representation of the radar echo trend during this motion, including both the overall motion trend and the trend of radar echo dissipation. Additionally, the \textit{TempEE} model exhibits a solid ability to represent a series of isolated points below.

Overall, the \textit{TempEE} model performs better than the Baseline model in terms of capturing the finer details of the radar echo distribution and the radar echo dissipation trend, making it a more suitable model for this application.

\noindent \textbf{Case Study on Spare-Nonstationary Motion Process.} Despite the sparse and non-stationary motion of the echo within a two-hour timeframe, the baseline model adequately characterizes the primary motion trend, which gradually moves towards the upper right, and the general shape of the echo, including the bar region on the right side. However, it cannot account for a considerable number of sparse echo features, including those that appear suddenly at T = 0.174. Additionally, the model predicts a significantly high echo intensity, indicating the need for improvement.

In contrast, the proposed \textit{TempEE} captures the primary motion trend in regions with dense echo features and can flexibly perceive and characterize the sudden appearance of sparse echo features during the extrapolation process. Although it also encounters issues with high predicted intensity, the overall effectiveness of the \textit{TempEE} significantly exceeds that of the baseline model.

\noindent \textbf{Case Study on Dense-Nonstationary Motion Process.}
In the presence of dense non-stationary motion and echoes occurring within 2 hours, baseline fails to capture a limited subset of echo features and exhibits a significantly lower prediction intensity. While PredRNN++ is capable of capturing the vast majority of these features, it lacks a refined feature representation. In contrast, the proposed \textit{TempEE} demonstrates superior performance in echo feature refinement, motion and echo generation, and perception of extinction trends. Furthermore, \textit{TempEE}'s robustness and superiority over the baseline model are evident in its ability to maintain efficacy despite the gradual dissipation of echo features over time.

The proposed \textit{TempEE} has multiple advantages for characterizing sparse echo features and perceiving the echo generation and dissipation processes in both stationary and non-stationary motion. Additionally, it can refine echo characteristics and capture internal details based on the echoes' overall shape. Most importantly, the extrapolation effect remains robust over time, ensuring reliable analysis of long-term data. Collectively, these features make \textit{TempEE} a valuable tool for studying echo characteristics in various settings.

\subsection{Ablation Study}
We conducted a series of ablation experiments to evaluate the necessity and effectiveness of the specific modules in our proposed \textit{TempEE}, including TE, SE, and MSTA. Both qualitative and quantitative analyses were conducted, and the overall results of the ablation experiments are summarized in Table III. Notably, the presence of conflicting MSTA and MHSA modules means that when MSTA is removed, MHSA is utilized as its attention module in the Decoder.

Our results demonstrate the indispensability of TE and Se for the Decoder, as we observe that they are consistently maintained alongside MSTA in all instances. Removing TE/SE results in a noteworthy performance decline, whereas removing MSTA while retaining TE/SE leads to some degradation in the performance of the proposed \textit{TempEE}. This observation implies that MSTA exhibits a robust capability for multi-scale feature characterization, underscoring the validity and necessity of TE, SE, and MSTA in the proposed \textit{TempEE}.

\begin{table}[H]
  \centering
  \caption{Performance of the ablation study in which \textit{w/o} means that the proposed \textit{\textit{TempEE}} without the certain module, \textit{w} means that conducts the experiment with the certain module, and $\uparrow$ and $\downarrow$ represent the higher/lower the metric, the better/poorer the performance.}
  \resizebox{0.8\textwidth}{!}{
    \begin{tabular}{ccccccccc}
    \toprule
    Model & TE & SE &  MSTA & MHSA & PNSR $\uparrow$  & SSIM $\uparrow$ & MSE $\downarrow$ & LPIPS $\downarrow$ \\
    \midrule
    \midrule
    \multirow{4}[1]{*}{\textit{\textit{TempEE}}} & \textit{w}  & \textit{w}  & \textit{w}  & \textit{w/o} & 32.43 & 0.876 & 29.82 & 1.058 \\
       & \textit{w/o} & \textit{w}  & \textit{w}  & \textit{w/o} & 29.42 & 0.750 & 53.53 & 1.687 \\
       & \textit{w}  & \textit{w/o} & \textit{w}  & \textit{w/o} & 30.10 & 0.767 & 46.65 & 1.556 \\
       & \textit{w}  & \textit{w}  & \textit{w/o} & \textit{w}  & 28.70 & 0.758 & 61.23 & 1.785 \\
    \bottomrule
    \end{tabular}}
  \label{tab:addlabel}
\end{table}%

The ablation experiment assessed the performance metrics of \textit{TempEE} based on image quality over time, as depicted in Fig. 18. The results demonstrated that all versions of \textit{TempEE} without TE, SE, and MSTA exhibited lower performance levels in comparison to the original \textit{TempEE}. However, the temporal trend remained generally stable, with the exception of a significant decay noticed at the final extrapolation time point caused by the random sampling strategy. These results further validate the superiority of the one-step forward extrapolation mechanism employed in the proposed \textit{TempEE}. This mechanism prevents the accumulation of errors during the extrapolation process and has the potential for ultra-long period extrapolation, which is unattainable with the auto-regressive extrapolation strategy.

\begin{figure}[H]
\centering
\includegraphics[width=0.95\textwidth]{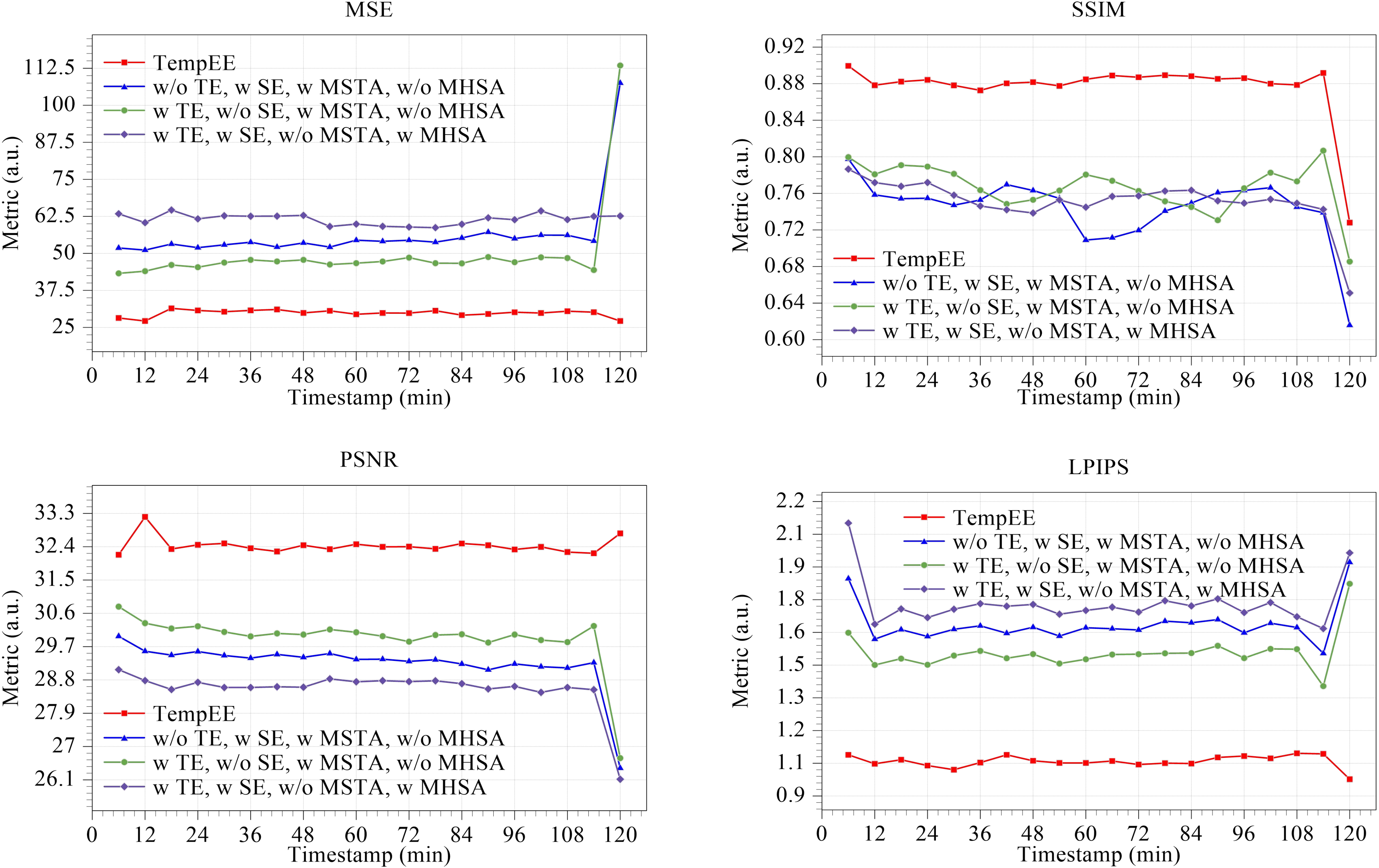}
\caption{Time-by-time performance comparison of the proposed \textit{TempEE} and \textit{TempEE}-based ablation model under image-based metrics.} 
\label{xas} 
\end{figure}

\begin{figure}[H]
\centering
\includegraphics[width=0.95\textwidth]{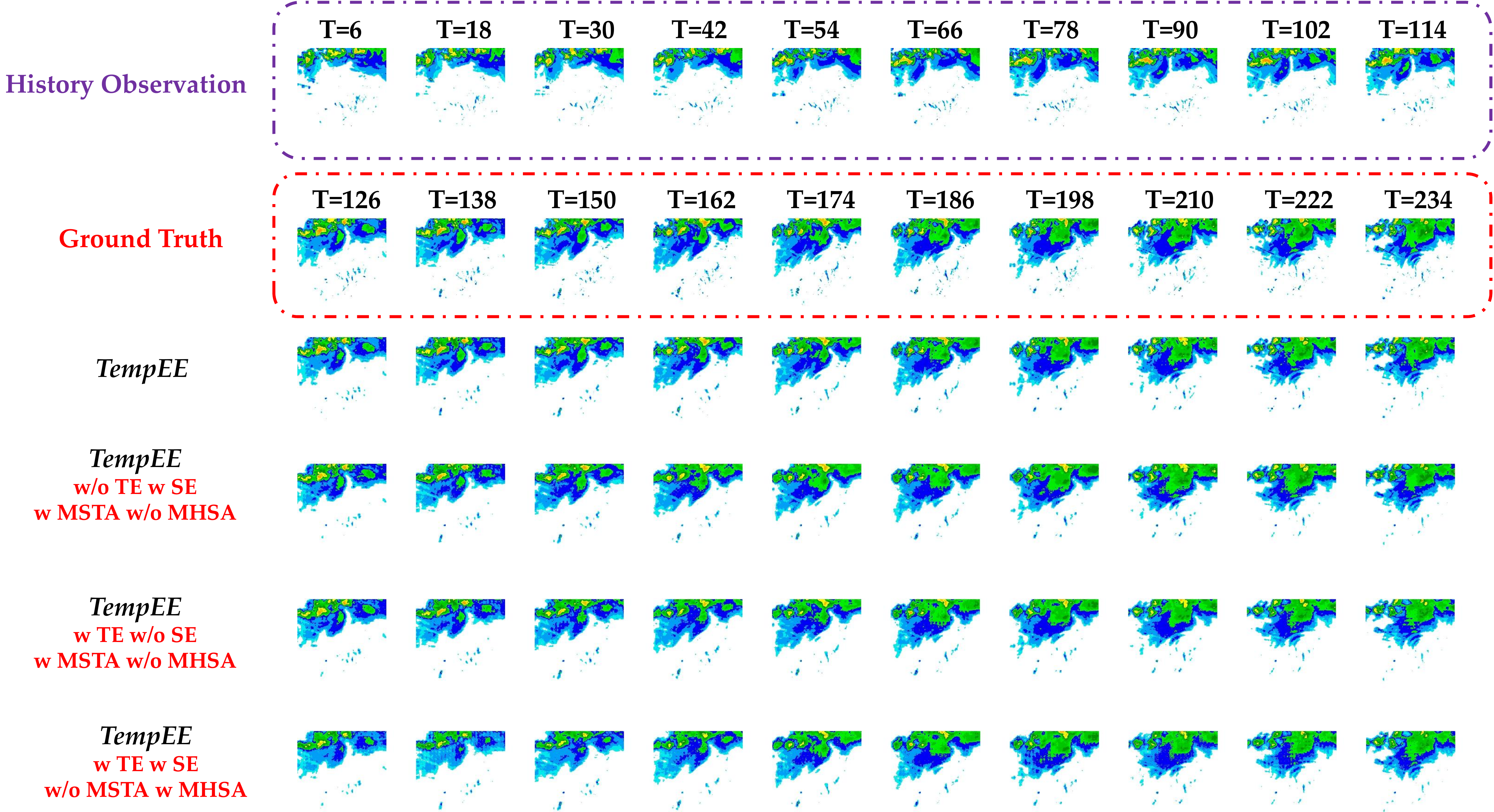}
\caption{Visualization of the the proposed \textit{TempEE} and its ablation model's inference process, note that the time period actually is 6 $min$, but 12 $min$ is used as the presentation period.} 
\label{ss} 
\end{figure}

The results of the ablation experiments are presented in Fig.~\ref{ss}, which depicts the dense and non-stationary process of echo motion with varying degrees of echo generation and extinction. The numerical comparison about performance is presented in Table IV. The results show that the original \textit{TempEE} has a significant impact when compared to the other ablation models. Furthermore, all models effectively capture the overall echo motion trend, sparse echo features, and abrupt changes in echo features. However, the ablation models show significant image gridding issues during the extrapolation process when compared to the original \textit{TempEE}. As a result, the predicted echo images are not clear enough, which could marginally affect the forecast results.

The proposed one-step forward extrapolation mechanism of \textit{TempEE} significantly outperforms the traditional auto-regressive extrapolation. It solves the fundamental issue of error accumulation in the extrapolation process of conventional extrapolation models. Furthermore, \textit{TempEE} consistently outperforms other models in cases of stationary, non-stationary, and different echo distributions.

\begin{table*}[b]
  \centering
  \caption{Results of the ablation study based on perception forecasting -based metrics, where $\tau$ is the evaluation threshold representing different rainfall intensities, and $\uparrow$ and $\downarrow$ represent the higher/lower the metric, the better/poorer the performance, and \textit{w} and \textit{w/o} denotes `with` and `without`, respectively.}
  \resizebox{1\textwidth}{!}{
    \begin{tabular}{ccccccccccccc}
    \toprule
    \multirow{2}[4]{*}{Model} & \multicolumn{6}{c}{CSI $\uparrow$}     & \multicolumn{6}{c}{ETS $\uparrow$} \\
\cmidrule{2-13}       & $\tau = 5$ & $\tau = 10$ & $\tau = 20$ & $\tau = 30$ & $\tau = 40$ & Average & $\tau = 5$ & $\tau = 10$ & $\tau = 20$ & $\tau = 30$ & $\tau = 40$ & Average \\
\midrule
\midrule
\textit{TempEE} & 0.861 & 0.849 & 0.719 & 0.588 & 0.54 & 0.711 & 0.822 & 0.746 & 0.699 & 0.579 & 0.534 & 0.676 \\
    \textit{w/o} TE \textit{w} SE \textit{w} MSTA \textit{w/o} MHSA & 0.794 & 0.773 & 0.655 & 0.48 & 0.422 & 0.625 & 0.74 & 0.734 & 0.625 & 0.461 & 0.406 & 0.593 \\
    \textit{w} TE \textit{w/o} SE \textit{w} MSTA \textit{w/o} MHSA & 0.812 & 0.802 & 0.678 & 0.519 & 0.464 & 0.655 & 0.763 & 0.768 & 0.648 & 0.501 & 0.449 & 0.626 \\
    \textit{w} TE \textit{w} SE \textit{w/o} MSTA \textit{w} MHSA & 0.782 & 0.758 & 0.604 & 0.382 & 0.32 & 0.569 & 0.727 & 0.709 & 0.573 & 0.365 & 0.306 & 0.536 \\
    \midrule
    \multirow{2}[4]{*}{Model} & \multicolumn{6}{c}{FAR $\downarrow$}     & \multicolumn{6}{c}{POD $\uparrow$} \\
\cmidrule{2-13}       & $\tau = 5$ & $\tau = 10$ & $\tau = 20$ & $\tau = 30$ & $\tau = 40$ & Average & $\tau = 5$ & $\tau = 10$ & $\tau = 20$ & $\tau = 30$ & $\tau = 40$ & Average \\
\midrule
\midrule
\textit{TempEE} & 0.045 & 0.088 & 0.158 & 0.231 & 0.273 & 0.159 & 0.934 & 0.882 & 0.828 & 0.716 & 0.682 & 0.808 \\
    \textit{w/o} TE \textit{w} SE \textit{w} MSTA \textit{w/o} MHSA & 0.77 & 0.149 & 0.186 & 0.271 & 0.315 & 0.200 & 0.913 & 0.841 & 0.756 & 0.571 & 0.513 & 0.719 \\
    \textit{w} TE \textit{w/o} SE \textit{w} MSTA \textit{w/o} MHSA & 0.069 & 0.135 & 0.178 & 0.242 & 0.281 & 0.181 & 0.921 & 0.856 & 0.780 & 0.608 & 0.554 & 0.744 \\
    \textit{w} TE \textit{w} SE \textit{w/o} MSTA \textit{w} MHSA & 0.072 & 0.144 & 0.150 & 0.220 & 0.255 & 0.168 & 0.892 & 0.799 & 0.661 & 0.414 & 0.349 & 0.623 \\
    \bottomrule
    \end{tabular}}
  \label{tab:addlabel}%
\end{table*}%

\section{Conclusion}
\label{sec_con}
In conclusion, this paper proposes a novel radar echo extrapolation model named \textit{TempEE}, providing reliable and precise analytical support for precipitation forecasting by extrapolating radar echoes. The proposed parallel encoder structure, consisting of a Temporal Encoder and a Spatial Encoder, along with a decoder that employs attention, considers complex spatiotemporal correlations among echo distributions simultaneously. Experimental results on the real-world radar echo dataset have demonstrated the effectiveness of \textit{TempEE} in achieving state-of-art performance with low computational cost. Moreover, extensive ablation studies confirm the necessity and efficacy of the model's components.

However, \textit{TempEE} necessitates substantial computational resources for calculating attention scores, which are relatively costly compared to convolution units. Furthermore, due to the usage of real-world datasets gathered from three radars with distinct location information, \textit{TempEE} neglects geographical association during the training phase. Our future work will focus on creating a foundational model for radar echo extrapolation tasks that incorporates geographical association while maintaining cost-effectiveness for broader applicability.

\section*{Acknowledgments}
This work was supported in part by the National Key Research and Development Program of China for Intergovernmental Cooperation under Grant 2019YFE0110100, in part by the National Natural Science Foundation of China under Grant 42105145, in part by the Guangdong Province Natural Science Foundation under Grant 2023A1515011438, in part by the Science and technology innovation team project of Guangdong Meteorological Bureau under Grant GRMCTD202104, in part by the Innovation and Development Project of China Meteorological Administration under Grant CXFZ2022J002, in part by the the Shenzhen Sustainable Development Project (No. KCXFZ20201221173610028), and in part by the Shenzhen Hong Kong Macao science and technology plan project under Grant SGDX20210823103537035.

\clearpage
\bibliographystyle{IEEEtran}
\bibliography{ref}{}
\end{document}